\documentclass[twocolumn]{aastex61}  % twocolumn, manuscript, linenumbers, tighten
\usepackage{amsmath}

\shorttitle{DISRUPTION OF LOW-MASS WDS WITH H ENVELOPES}
\shortauthors{LAW-SMITH ET AL.}

\begin{document}

\title{LOW-MASS WHITE DWARFS WITH HYDROGEN ENVELOPES\\ AS A MISSING LINK IN THE TIDAL DISRUPTION MENU}

\correspondingauthor{Jamie Law-Smith}
\email{lawsmith@ucsc.edu}

\author{Jamie Law-Smith}
\affiliation{Department of Astronomy and Astrophysics, University of California, Santa Cruz, CA 95064, USA}

\author{Morgan MacLeod}
\altaffiliation{NASA Einstein Fellow}
\affiliation{School of Natural Sciences, Institute for Advanced Study, 1 Einstein Drive, Princeton, NJ 08540, USA}

\author{James Guillochon}
\affiliation{Harvard-Smithsonian Center for Astrophysics, The Institute for Theory and Computation, 60 Garden Street, Cambridge, MA 02138, USA}

\author{Phillip Macias}
\affiliation{Department of Astronomy and Astrophysics, University of California, Santa Cruz, CA 95064, USA}

\author{Enrico Ramirez-Ruiz}
\affiliation{Department of Astronomy and Astrophysics, University of California, Santa Cruz, CA 95064, USA}

\begin{abstract}
We construct a menu of objects that can give rise to bright flares when disrupted by massive black holes (BHs), ranging from planets to evolved stars. Through their tidal disruption, main sequence and evolved stars can effectively probe the existence of otherwise quiescent supermassive BHs and white dwarfs can probe intermediate mass BHs. Many low-mass white dwarfs possess extended hydrogen envelopes, which allow for the production of prompt flares in disruptive encounters with moderately massive BHs of $10^5$--$10^7~M_\sun$---masses that may constitute the majority of massive BHs by number. These objects are a missing link in two ways: (1) for probing moderately massive BHs and (2) for understanding the hydrodynamics of the disruption of objects with tenuous envelopes. A flare arising from the tidal disruption of a $0.17~M_\sun$ white dwarf by a $10^5~M_\sun$ BH reaches a maximum between 0.6 and 11 days, with a peak fallback rate that is usually super-Eddington and results in a flare that is likely brighter than a typical tidal disruption event. Encounters stripping only the envelope can provide hydrogen-only fallback, while encounters disrupting the core evolve from H- to He-rich fallback. While most tidal disruption candidates observed thus far are consistent with the disruptions of main sequence stars, the rapid timescales of nuclear transients such as Dougie and PTF10iya are naturally explained by the disruption of low-mass white dwarfs. As the number of observed flares continues to increase, the menu presented here will be essential for characterizing nuclear BHs and their environments through tidal disruptions.
\end{abstract}


% ----------------------------------------------------------------------------
\section{INTRODUCTION}

When a star wanders too close to a massive black hole (MBH), it can be ripped apart by the hole's tidal field \citep{1975Natur.254..295H, 1978MNRAS.184...87F, 1979Natur.280..214G, 1982Natur.296..211C, 1988Natur.333..523R}. In a typical disruption, half of the material will be ejected on hyperbolic trajectories and half of the material will remain bound to the MBH; the accretion of this material gives rise to a transient usually referred to as a tidal disruption event (TDE). With accurate theoretical modeling, TDEs allow us to uncover the mass of the black hole, the characteristics of the surrounding stellar population, the dynamics of the galactic nucleus, and the physics of black hole accretion under well-defined conditions \citep{2014ApJ...783...23G}. TDEs can also provide a direct and unambiguous probe of the MBH occupation fraction of low-mass galaxies, which is crucial for constraining MBH seed formation efficiency at high redshifts---a dominant mechanism of initial galaxy formation \citep{2012ApJ...760..103D, 2015ApJ...809..166G, 2016MNRAS.455..859S, 2016MNRAS.461..371K}. The opportunity to study BHs in the local universe through TDEs is important, because for every actively accreting BH, there are $\sim 170$ quiescent BHs \citep{2007ApJ...667..131G,2010PhRvD..81j4014G}.

TDEs are observationally identified by a combination of a dramatic increase in brightness, proximity to a non-active host galaxy's center, and weak or no color evolution at optical/UV wavelengths, with a decay in luminosity that is theoretically predicted to follow a $t^{-5/3}$ law \citep[for reviews of the observations, see e.g.][]{2015JHEAp...7..148K,2017ApJ...838..149A}. The most compelling events are those in which the rise, peak, and decay of the optical/UV transient are observed with frequent cadence, as each of these phases of a TDE contain vital information about the disruption, and can be used to constrain the properties of the host black hole and the object that was disrupted \citep[e.g.,][]{2012Natur.485..217G, 2014ApJ...783...23G}. Taken in a statistical sense, the observed rates of tidal disruption and, in particular, the relative rates of disruptions of different stellar objects, will hold tremendous distinguishing power in terms of both the dynamical mechanisms operating in galactic centers and the properties of the populations of stars themselves \citep{2012ApJ...757..134M,2014ApJ...794....9M,2016ApJ...819....3M}.

A central objective of this work is to understand the menu of all possible TDEs about massive BHs---i.e., which objects produce tidal disruption flares for which BH masses, and how they dictate the properties of the fallback accretion rate onto the BH. An object of mass $M$ and radius $R$ can be torn apart if it crosses the tidal radius, $r_\text{t} = (M_\text{bh}/M)^{1/3}\ R$, of a BH with mass $M_\text{bh}$. Therefore, the characteristics of a particular stellar object hold information about the nature of its disruption---whether it occurs near the BH's innermost bound circular orbit, and, if so, how relativistic the encounter is. BHs with masses $\gtrsim 10^7~M_\sun$ are well probed by MS stars, evolved stars, and planets, but the debris could be ineffective at circularizing for BHs with masses $\lesssim 10^6~M_\sun$, as shown by semi-analytic results in \citet{2015ApJ...809..166G}, as well as the Newtonian and relativistic hydrodynamic simulations of \citet{2014ApJ...783...23G} and \citet{2015ApJ...804...85S}. BHs with masses $\lesssim 10^5~M_\sun$ could be probed by typical white dwarfs \citep[although BH spin could raise this limit to $10^6~M_\sun$;][]{2017arXiv170100303T}.

Thus far, most observed TDE candidates come from host galaxies with inferred BH masses of $\gtrsim 10^6~M_\sun$. Even though survey selection effects make seeing TDEs from lower-mass BHs less likely \citep[see e.g.,][]{2016MNRAS.461..371K}, we should expect to observe them with future surveys if the BH mass function is not truncated below $10^6~M_\sun$. Tidal disruption flares are potentially a powerful probe of the galaxy occupation fraction of these BHs, and could help discriminate between BH mass functions that are flat, rising (as extrapolated from the $M$--$\sigma$ relation), and/or truncated at low masses. Our ability to use TDEs as direct probes of black hole demographics necessitates a detailed understanding of how the observability of TDEs depends on the properties of the disrupted star. Constructing a complete menu of stellar tidal disruption simulations---as we do in this work---is an important step in addressing these questions.

Theoretical studies of stellar structure and fallback rate began with Lagrangian \citep{1989ApJ...346L..13E} and Eulerian \citep{1993ApJ...418..163K} calculations, and have evolved to include detailed studies of MS stars \citep{2009MNRAS.392..332L, 2009ApJ...697L..77R, 2013ApJ...767...25G}, giant planets \citep{2011ApJ...732...74G, 2013ApJ...762...37L}, white dwarfs \citep{1989A&A...209..103L, 2004ApJ...615..855K, 2008ApJ...679.1385R, 2008CoPhC.179..184R, 2009ApJ...695..404R, 2010MNRAS.409L..25Z, 2011ApJ...726...34C, 2011ApJ...743..134K, 2012ApJ...749..117H, 2014PhRvD..90f4020C, 2014ApJ...794....9M, 2016ApJ...819....3M, 2016arXiv161207316V}, and giant stars \citep{2012ApJ...757..134M, 2013ApJ...777..133M}. 

A finding common to all calculations is that a more centrally concentrated object has a quicker-peaking fallback rate and requires a deeper encounter for full disruption than a less centrally concentrated object. Here, ``deeper" is in relation to the tidal radius definition, which relates to the average density. The presence of a core is also important in determining the fallback rate; in giant stars, the massive core plays a key role yet typically remains intact, while in giant planets, the lighter core is much more vulnerable. These considerations are crucial, as we expect the stellar structure to be imprinted on the luminosity evolution of the flare. In many of the observed events, the luminosity evolution closely follows the predicted mass fallback onto the BH \citep[a classic example is PS1-10jh;][]{2012Natur.485..217G, 2014ApJ...783...23G}. This preservation of the fallback rate implies that circularization of the debris is prompt in these cases; the mass feeding rate is primarily determined by fallback and is not significantly delayed by viscous effects. 

Flares can be delayed if the amount of energy dissipated per orbit---or ``viscosity''---is small. When the stream's self-intersection point is relatively close to the BH, energy dissipation is large, allowing the debris to circularize quickly \citep{2016MNRAS.455.2253B}. Once the disk is formed, the viscous transport timescale (i.e., the time it takes material to accrete) at the circularization radius is much shorter than the peak fallback timescale. When the stream's self-intersection point is much farther from the BH than the periapse distance, however, circularization is not effective, and a highly elliptical disk is formed \citep{2009ApJ...697L..77R, 2015ApJ...804...85S}. In this case, the viscous timescale can be significantly longer than the peak fallback timescale \citep{2015ApJ...809..166G}.

Stellar structure in tidal disruption calculations has thus far been implemented using polytropic profiles, with the simplest examples being the single-polytrope models of MS stars and WDs. Evolved stars and planets with cores are not well described by a single polytrope; these objects have been studied using a nested polytrope in which the envelope is a significant fraction of the total mass \citep{2012ApJ...757..134M, 2013ApJ...762...37L}. 

In this work we perform the first tidal disruption calculations for objects where the atmosphere has a small mass relative to the core, with our primary motivating physical example being a low-mass He WD with a hydrogen envelope---though we note that this structure could potentially also be used to model hot Jupiters or very evolved stars. Any WD below $\approx 0.46~M_\sun$ has a helium core, and possesses a hydrogen envelope that, despite its comparatively low mass, can extend to several times the core's radius \citep[e.g.,][]{2001A&A...365..491N}. In this work, we calculate the disruption of these objects and predict their observational properties. We argue that these objects are a missing link in two ways: (1) for probing moderately massive BHs, and (2) for understanding the hydrodynamics of the disruption of objects with tenuous envelopes, as such structures have not yet been studied. We find that these low-mass WDs with hydrogen envelopes offer prompt flares at higher-mass BHs than their more typical WD counterparts, and occupy a unique parameter space in time and luminosity at peak.

In Section \ref{sec:menu}, we develop the tidal disruption menu, which is our motivation for the hydrodynamical simulations of this paper. In Section \ref{sec:hwds}, we discuss the particulars of He WDs. In Section \ref{sec:setup}, we outline our hydrodynamical setup for disrupting these objects, and in Section \ref{sec:results} we present numerical results from these simulations. In Section \ref{sec:demographics}, we present an overview of tidal disruption flare demographics in terms of peak timescales and fallback rates. In Section \ref{sec:discussion}, we summarize our findings and show that fast-rising events such as Dougie and PTF10iya are naturally explained by the disruption of an He WD.

% ----------------------------------------------------------------------------
\section{TIDAL DISRUPTION MENU}\label{sec:menu}

To determine whether an object is disrupted or swallowed by a black hole, we need to compare the tidal radius, $r_\text{t}$, to the innermost bound circular orbit of the black hole,
\begin{equation}
r_\text{ibco} = \frac{2GM_\text{bh}}{c^2} \left(1 - \frac{a_\ast}{2} + \sqrt{1-a_\ast} \right),
\end{equation}
where $a_\ast = a/M$, $a = J_\ast/M_\ast c$, $M = GM_\ast / c^2$, and $M_\ast$ and $J_\ast$ are the mass and angular momentum of the BH, respectively \citep{2013LRR....16....1A}. For a non-spinning BH, $r_\text{ibco}=4GM_\text{bh}/c^2$, and for a maximally spinning BH, $r_\text{ibco}=GM/c^2$. If $r_\text{t} > r_\text{ibco}$, disruption is possible. Otherwise, the object is swallowed whole \citep[e.g.,][]{2014ApJ...795..135E}. For simplicity we assume here that disruption is only possible when the impact parameter $\beta = r_\text{t}/r_\text{peri} \geq 1$; more accurately, disruption is a smooth function of $\beta$. For a non-spinning BH, we therefore require
\begin{equation}
M_\text{bh} \le M_\text{bh, lim} = \frac{R_\star^{3/2}}{M_\star^{1/2}} \left(\frac{c^2}{4G}\right)^{3/2} \propto \rho_\star^{-1/2}
\end{equation}
for disruption. The mass-radius relationship, then, determines whether an object will be disrupted at a given BH mass. Denser objects such as WDs can only be disrupted by lower-mass BHs while more tenuous objects such as MS or evolved stars can be disrupted by higher-mass BHs.

\begin{figure}[tbp]
\epsscale{1.22}
\plotone{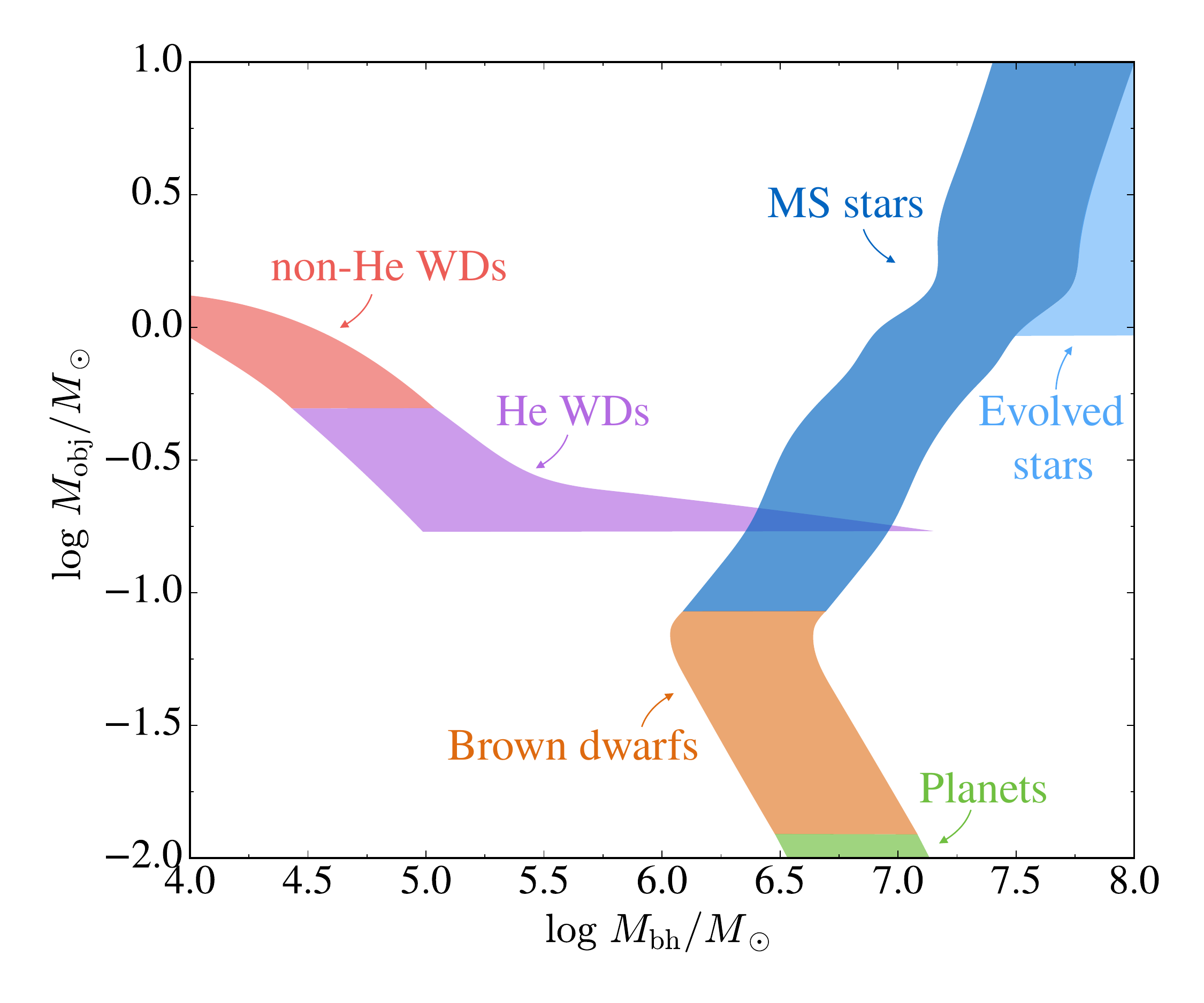}
\caption{
Regions where prompt tidal disruption flares are favorable in $M_\text{obj}$ vs. $M_\text{bh}$ space for a non-spinning BH. Encounters have $4GM_\text{bh}/c^2 < r_\text{t} < 10 GM_\text{bh}/c^2$. Note that disruption is still possible for lower BH masses than shown in each region. We include mass-radius relationships for typical WDs from \citet{2010MNRAS.409L..25Z}, MS stars from \citet{1996MNRAS.281..257T}, evolved stars from \citet{2012MNRAS.427..127B, 2013EPJWC..4303001B}, and sub-stellar objects from the 1 Gyr curve of \citet{2009AIPC.1094..102C}. We define MS stars as $M \geq 0.085~M_\sun$, brown dwarfs as $0.085~M_\sun > M \geq 13~M_\text{Jup}$, and planets as $M < 13~M_\text{Jup}$. For evolved stars, we choose masses above $0.9~M_\sun$ (here the evolutionary time is approximately equal to the Hubble time) and radii up to the radius at the tip of the red giant branch for this mass star. WDs below $\sim 0.5~M_\sun$ will be helium-core hydrogen-envelope WDs. We calculate the radii of three He WDs 1 Gyr after formation and interpolate for masses in between. We choose representative masses of $0.17~M_\sun$, $0.25~M_\sun$, and $0.38~M_\sun$, with initial envelope masses of $0.011~M_\sun$, $0.016~M_\sun$, and $0.019~M_\sun$ respectively. This is motivated by the fact that the mass distribution of He WDs is expected to be relatively flat \citep{2012ApJ...751..143M}.
}
\label{menu}
\end{figure}

We can calculate the upper limit for the disruption of a class of objects by using the above relation. We show this menu of BH-object combinations for a non-spinning BH, along with a prompt circularization condition explained below, in Figure \ref{menu}. We use mass-radius relations for WDs, MS stars, evolved stars, and sub-stellar objects. We find that He WDs with hydrogen envelopes play a special role in this menu, as, similar to evolved stars, they can have a wide range of radii at a given mass, depending on their age. Compared to the relatively tight mass-radius relation for typical white dwarfs, these objects allow access to a higher range of BH masses. More details on He WDs and our stellar evolution calculations of their structure are given in Section \ref{sec:hwds}. Our choice of representative masses and ages is justified there.

Many of the tidal disruption candidates observed thus far show a luminosity time evolution that closely follows the mass fallback rate from the star to the BH \citep[see e.g.,][]{2014ApJ...783...23G}. This suggests that current observations may select for events in which debris circularization is prompt. Recent work suggests that prompt circularization occurs predominantly for encounters where general relativistic effects are important \citep{2013MNRAS.434..909H, 2015ApJ...812L..39D, 2015ApJ...809..166G}. We take a ``circularization condition'' of $r_\text{t} < 10 GM_\text{bh}/c^2$ in order to select encounters in this regime. Following \citet{2015ApJ...812L..39D} and \citet{2015ApJ...809..166G}, this corresponds to a de Sitter apsidal precession of $\Omega \gtrsim 54^\circ$ for non-spinning BHs. Note that more weakly plunging encounters will still circularize some fraction of the time, and that they may also be observable as events where the luminosity evolution is viscously delayed; our condition is meant as a guideline for where we can expect to see predominantly prompt circularization events for a given disruptee. Note also that most events in the X-rays appear to be viscously delayed \citep{2017ApJ...838..149A}.

For a non-spinning BH, our condition for prompt flares is then $4GM_\text{bh}/c^2 < r_\text{t} < 10 GM_\text{bh}/c^2$. WDs can only be disrupted by BHs with masses $\lesssim 10^5~M_\sun$, while MS and evolved stars only obey our prompt flare condition for BHs with masses $\gtrsim 10^6~M_\sun$. Because of their extended radius, low-mass WDs with hydrogen envelopes can serve as a missing link between these two regimes. Their envelope can be disrupted and stripped by higher BH masses than allowed for by typical WDs. These BH masses offer a relatively smaller fraction of prompt flares from MS stars due to their inefficient circularization here. The constraints derived for He WDs, here assumed to be at least 1 Gyr after formation, could be extended to higher-mass BHs for younger He WDs, which have significantly more extended envelopes. For example, a 100 Myr old $0.17~M_\sun$ He WD can have a radius of $0.5~R_\sun$, allowing it to be disrupted by a $10^8~M_\sun$ BH.

Low-mass WDs can thus extend the range of BH masses available to the higher-mass, single-star evolution WDs through tidal disruption.\footnote{There is some evidence now mounting for observational candidates of WD disruptions by intermediate mass BHs. In particular, an emerging class of ultra-long gamma-ray burst (ULGRB) sources share similar timescales and luminosities to WD disruptions \citep[see][]{2014ApJ...781...13L, 2015JHEAp...7...44L, 2014ApJ...794....9M, 2016ApJ...819....3M}.} While these objects make up a small fraction of the stellar population, they deserve to be examined in more detail because of their unique location in our prompt circularization menu, which, as we argue, makes their emerging flares more favorable to detection.

%-----------------------------------------------------------------------------
\section{Helium-core Hydrogen-envelope WDs}\label{sec:hwds}

\subsection{Properties}\label{subsec:properties}

\begin{figure*}[tbp]
\epsscale{1.1}
\plottwo{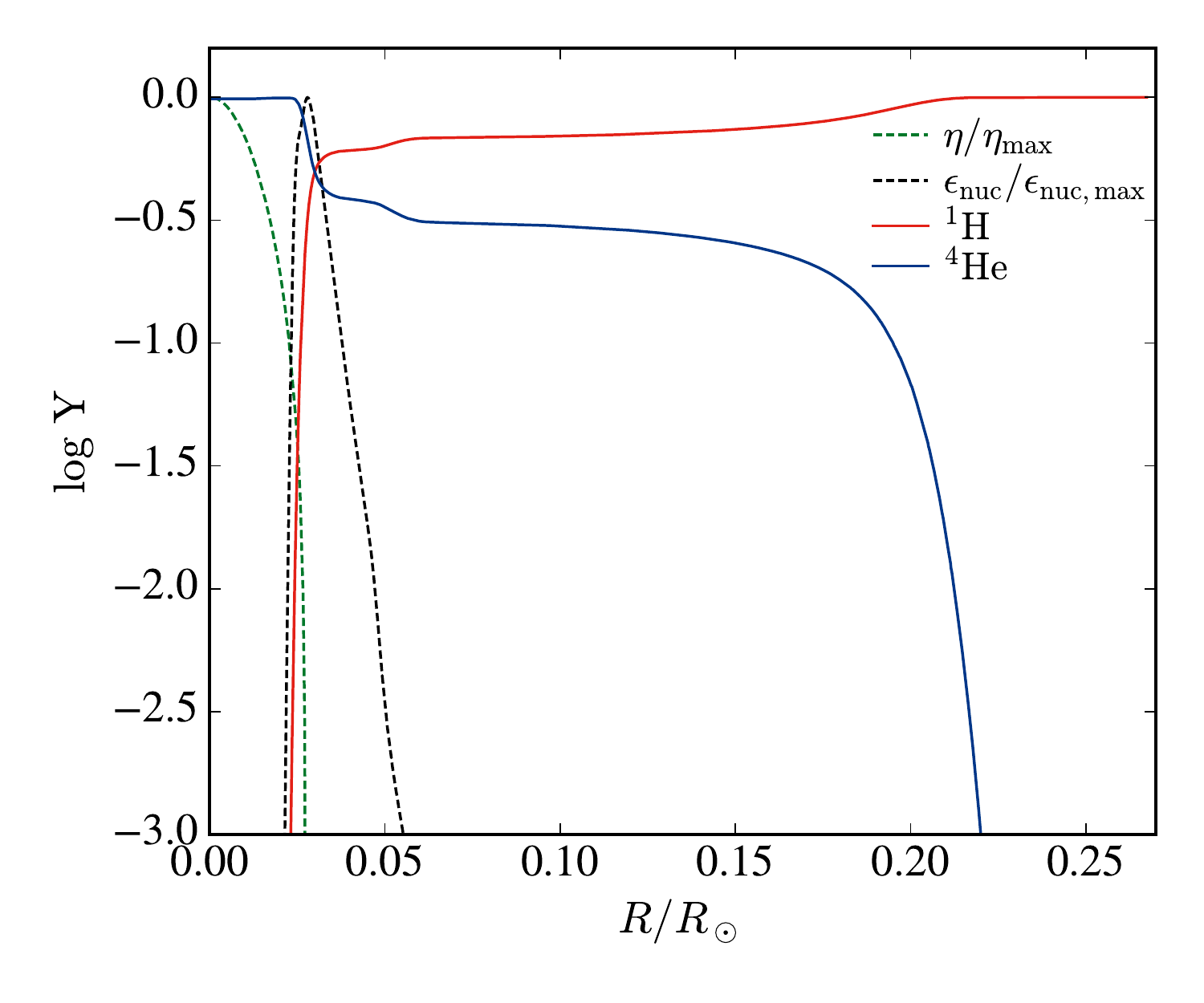}{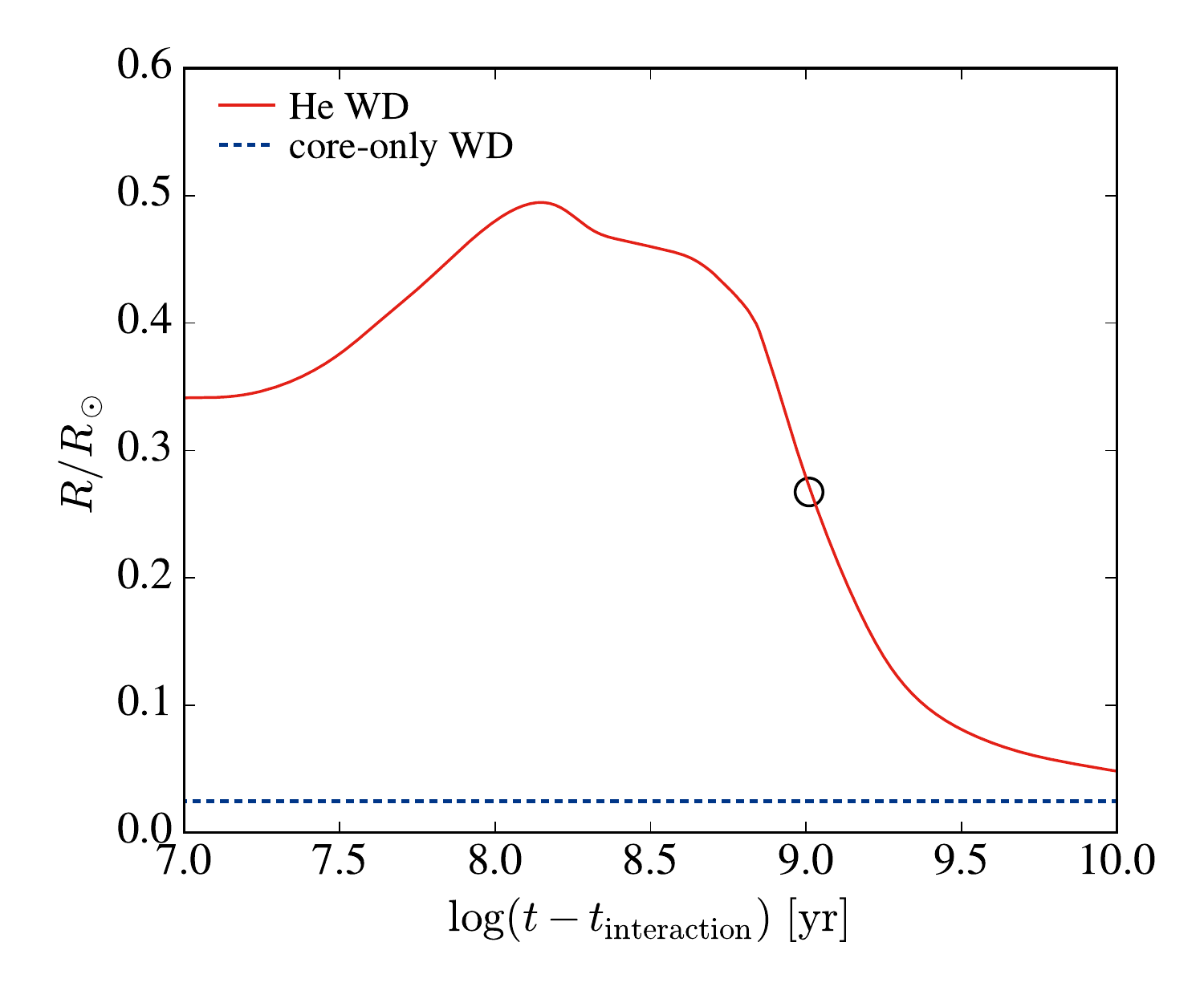}
\caption{
Left panel: helium (blue) and hydrogen (red) abundances as a function of radius for a $0.17~M_\sun$ helium-core hydrogen-envelope WD 1 Gyr after formation. The mass of the hydrogen envelope is only $10^{-2}~M_\sun$, but it extends to roughly 10 times the radius of the core. The green dashed line shows the degeneracy parameter $\eta$, indicating the degenerate helium core. The black dashed line shows the nuclear burning fraction $\epsilon_\text{nuc}$, indicating the thin hydrogen burning region surrounding the core. $\eta$ and $\epsilon_\text{nuc}$ are shown normalized to their maximum values. Right panel: radius as a function of time since formation (through a binary interaction) for this WD. Its radius is 10 times larger than that of a WD without an envelope, shown in dashed blue, for more than 1 Gyr. The black circle indicates the age and radius of the object we use in our disruption calculations.
}
\label{abundances_burning}
\end{figure*}

Since WDs have an inverse mass-radius relationship, the lowest mass WDs will be able to probe the highest mass BHs. Let us estimate the lowest mass WD available through single-star evolution. Setting the main sequence lifetime equal to the age of the universe \citep[$\approx 13.8~\text{Gyr}$;][]{2000MNRAS.315..543H} using an analytic formula for the MS lifetime from \citet{2013ApJS..208...19H} gives $M_\text{i} \approx 0.9~M_\sun$. Using this mass in an empirical initial--final mass relation for WDs from \citet{2008MNRAS.387.1693C} for $M_\text{i} < 2.7~M_\sun$,
\begin{equation}
M_\text{f} = (0.096 \pm 0.005) M_\text{i} + (0.429 \pm 0.015),
\end{equation}
we find that the minimum WD mass possible through single-star evolution is $M_\text{WD} \approx 0.5~M_\sun$. 

WDs less massive than roughly half a solar mass will have formed through binary interactions, barring cases of extreme metallicity \citep{2007ApJ...671..761K}. Low-mass WDs can be formed either through stable Roche-lobe overflow mass transfer or common-envelope evolution \citep[e.g.,][]{1998A&A...339..123D,2000MNRAS.316...84S,2004ApJ...616.1124N,2013A&A...557A..19A,2015MNRAS.450L..39N}. A helium-core WD forms if one component of the binary loses its hydrogen envelope before helium burning. This object has a degenerate helium core and is formed with an extended hydrogen envelope supported by a thin hydrogen burning layer.

The maximum mass of an He WD is approximately $0.46~M_\sun$, and only He WDs are formed below this mass \citep{1990ApJ...364..527S,2001A&A...365..491N}. The final mass of the He WD depends on the mass of the progenitor and the binary orbital properties \citep[e.g.,][]{2001A&A...365..491N}. The progenitor star needs a zero-age main sequence mass below $2.3~M_\sun$, as more massive stars do not form helium cores. The strict minimum timescale for formation of an He WD is therefore the MS lifetime of a $2.3~M_\sun$ star, $t_\text{MS} \approx 1.16~\text{Gyr}$ \citep{2000MNRAS.315..543H}.

\citet{2014A&A...571L...3I,2016A&A...595A..35I} performed calculations of He WD formation via stable mass transfer; we quote some results below. After detachment from Roche-lobe overflow, the progenitor star enters a ``bloated'' proto-WD phase where much of the hydrogen in the envelope is burned in stable hydrogen shell burning. The mass of hydrogen left after Roche-lobe detachment is on the order of $10^{-2}~M_\sun$, yet this can fuel a proto-WD phase lasting up to 2.5 Gyr for the lowest mass ($M \lesssim 0.20~M_\sun$) WDs.

\citet{2014A&A...571L...3I} derived a timescale for hydrogen burning,
\begin{equation}\label{eq:deltatproto}
\Delta t_\text{proto} \simeq 400\ \text{Myr} \left(\frac{0.20~M_\sun}{M_\text{WD}}\right)^7,
\end{equation}
which describes the star's contraction from Roche-lobe detachment to its maximum effective temperature on the cooling track. \citet{2016A&A...595A..35I} also defined a cooling timescale, $t_\mathrm{cool, L_{-2}}$, which is the time from detachment to reaching $\log (L/L_\sun)=-2$ on the cooling track. This timescale is set primarily by the mass of the hydrogen envelope left at the end of the proto-WD phase. Generally, a shorter orbital period at the onset of mass transfer leads to a lower proto-WD mass and a higher final envelope mass.

There is a growing body of observations of these low-mass objects: the targeted survey for extremely low-mass (ELM; $M < 0.3~M_\sun$) WDs has found 76 binaries to date, with a median primary mass of $\approx 0.18~M_\sun$ \citep{2016ApJ...818..155B, 2016ApJ...824...46B}. Many of these WDs appear to be bloated, and this bloated state can persist for a long time: \citet{2015arXiv150400007M} find that roughly half of these systems will still be burning hydrogen when they merge. One object in this sample is the binary system NLTT 11748 \citep{2014ApJ...780..167K}, which contains a helium-core hydrogen-envelope WD of mass $0.17~M_\sun$ and radius $0.043~R_\sun$, whereas a standard WD mass-radius relation for this mass would give a radius of $0.02~R_\sun$. This object's bloated size allows it to be disrupted by a BH of up to $3.8 \times 10^6~M_\sun$. This WD has a cooling age of 1.6--1.7 Gyr; younger He WDs can have much more extended envelopes, allowing them to be disrupted by even $10^7~M_\sun$ or $10^8~M_\sun$ BHs. As an example of this more extreme bloating, observations and astroseismological studies of the eclipsing binary J0247--25 find a He WD with mass $0.186 \pm 0.002~M_\sun$ and radius $0.368 \pm 0.005~R_\sun$ \citep{2013Natur.498..463M}. Note that this He WD has a larger radius than a MS star of its mass. In a study of the Galactic WD binary population, \citet{2012ApJ...751..143M} found that roughly half of WDs in binaries are He WDs, and that the probability density distribution for He WDs is relatively flat below $0.4~M_\sun$.

It is difficult to estimate the typical age of a He WD upon disruption by a MBH, as these objects are formed from a range of progenitor masses and undergo a binary interaction of uncertain timescale. We do know that nuclear star clusters exhibit a wide range of stellar ages. For example, observations of the nearby S0 galaxy NGC 404 show that half of the mass of the nuclear star cluster is from stars with ages of $\approx1$ Gyr, while the bulge is dominated by much older stars \citep{2010ApJ...714..713S}. In our own Galactic center, roughly 80\% of the stars formed over 5 Gyr ago and the remaining 20\% formed in the last 0.1 Gyr \citep{2011ApJ...741..108P}. In addition, TDEs have so far been found preferentially in post-starburst galaxies, with significant 1 Gyr old or younger stellar populations \citep{2014ApJ...793...38A, 2016ApJ...818L..21F}. Another consideration is that in a study of a population of He WDs in the globular cluster NGC 6397, \citet{2003ApJ...586.1364H} found that the progenitor binaries of the He WDs very likely underwent an exchange interaction within the last Gyr. Finally, we note that the two-body relaxation time is $\approx 0.1$ Gyr for a $10^5~M_\sun$ BH and $\approx 1.8$ Gyr for a $10^6~M_\sun$ BH. Motivated by the above considerations, in our disruption simulations we take the radius of the He WD at 1 Gyr after formation (i.e., since Roche-lobe detachment). 

For the tidal disruption calculations in this work, we construct a $0.17~M_\sun$ He WD consisting of a $0.16~M_\sun$ degenerate helium core and a $0.01~M_\sun$ hydrogen envelope using the MESA stellar evolution code \citep{2011ApJS..192....3P, 2013ApJS..208....4P, 2015ApJS..220...15P}. This envelope mass is consistent with theoretical predictions of hydrogen retention \citep{2001MNRAS.323..471A, 2001MNRAS.325..607S, 2007MNRAS.382..779P}. The left panel in Figure \ref{abundances_burning} shows the relative abundance of helium and hydrogen as a function of radius for this object. The hydrogen envelope extends to roughly 10 times the radius of the core, and is supported by a thin hydrogen burning shell. This snapshot is at 1 Gyr after formation.

We also calculate the radius as a function of time since formation (through a binary interaction) for several He WDs in MESA. This is shown for our $0.17~M_\sun$ object in the right panel of Figure \ref{abundances_burning}. We show the radius of a core-only WD of the same mass for comparison in dashed blue (here we show a fixed radius that does not evolve with time). In a similar calculation for a $0.15~M_\sun$ WD, we find that a very extended envelope persists for $>10$ Gyr.

\subsection{Disruption and Flaring Rates}\label{subsec:rates}

The particular tidal disruption rates of different types of objects depend on the detailed dynamics and evolution of the dense stellar system surrounding the central BH. Given these uncertainties, here we make a simple estimate of the relative rate of He WD disruptions. We find that several factors could increase the rate from that suggested by these objects' low population fraction. We can decompose the observed rate into (1) the fractional disruption rate and (2) the rate of luminous flares.

\subsubsection{Disruption}
First, the fractional disruption rate. This can be written as $f_\text{disrupted} = f_\text{pop} \times f_\text{rel}$, where $f_\text{pop}$ is the fraction of the stellar population that are He WDs, and $f_\text{rel}$ is the specific likelihood of an He WD being disrupted. First we estimate $f_\text{pop}$. Modeling the Galactic population of double WDs, \citet{2001A&A...365..491N} found a Galactic birth rate of close double white dwarfs of $0.05~\text{yr}^{-1}$ and a formation rate of planetary nebulae of $1~\text{yr}^{-1}$. They found that $63\%$ of the stars in these pairs are He WDs. This implies that the production rate of He WDs is approximately $0.05 \times 0.63 \approx 0.03$ times that of single stellar evolution WDs. Choosing an age of 10 Gyr for the Galactic disk gives a turnoff mass of approximately $1~M_\sun$ for the stars in our Galaxy. We estimate the WD fraction by dividing the number of stars with masses of 1--$8~M_\sun$ (those that evolve to leave WD remnants) by the number with masses of 0.1--$8~M_\sun$ using a \citet{2001MNRAS.322..231K} IMF; this gives a WD fraction of approximately 0.16. The population fraction of He WDs is then $f_\text{pop} \approx 0.16 \times 0.03 \approx 0.005$.

There is a concern that mass segregation might limit $f_{\rm pop}$ in central cluster regions. In clusters, low-mass stars are evaporated from the central regions as above-average mass objects settle deeper in a trend toward energy equipartition on the cluster relaxation time \citep[e.g.,][]{2013degn.book.....M}. However, binaries containing an He WD, even though the He WD mass is low, will not be evaporated from the central regions as their total mass is on average higher than the average mass of a typical stellar population. Indeed, in a study of the central regions of globular cluster NGC 6397, \citet{2009ApJ...699...40S} found a sample of He WDs with masses of 0.2--$0.3~M_\sun$. These objects show strong H$\alpha$ absorption lines (indicating that they still retain their hydrogen envelopes), and are significantly more concentrated in the cluster center than either the CO WDs or the turnoff stars. We therefore expect that mass segregation either enhances $f_{\rm pop}$ or, at least, does not reduce it in nuclear star clusters.

This population fraction could also be larger due to the fact that in dense stellar systems, the rate of dynamically assembled compact binaries is observed to be enhanced by a factor of 10--100 when compared to the field \citep{2003ApJ...591L.131P, 2006ApJ...646L.143P}. We might expect similar enhancements in the dense and dynamical nuclear region surrounding an MBH. Note that the separation of He WDs from their companions is observed to be $10^{10}~\text{cm} < a < 3 \times 10^{11}~\text{cm}$ in the ELM survey, making these binaries stable against ionization for typical nuclear cluster conditions.

For $f_\text{rel}$, we follow \citet{2012ApJ...757..134M} and scale the specific likelihood of disruption as $f_\text{rel} \propto r_\text{t}^{1/4}$. Relative to an MS star, this is $f_\text{rel} = (R_\text{He}/R_\text{MS})^{1/4}(M_\text{MS}/M_\text{He})^{1/12}$, which is of order unity for our $0.17~M_\sun$ He WD and a $\sim 0.5~M_\sun$ MS star. This gives us a conservative total fractional disruption rate of $f_\text{disrupted} = f_\text{pop} \times f_\text{rel} \approx 0.005$. For a hydrogen-depleted He WD, $f_\text{rel}$ is closer to $1/2$. As a potential comparison, simulations of star clusters by \citet{2004ApJ...613.1143B} found that the relative fraction of WD disruptions is $\approx0.15$. Multiplying this by the relative production rate of He WDs \citep[roughly 0.03 for every WD in our Galaxy following][]{2001A&A...365..491N} suggests a fractional disruption rate of $f_\text{disrupted} \approx 0.005$, consistent with our above estimate. However, as mentioned, mass segregation and dynamical assembly effects can enhance our above estimate. The estimate using star cluster simulations may also be low, as these simulations include very low-mass BHs and a population of single stars. These calculations therefore model the disruption of only single WDs, which also follow the substantially more compact typical WD mass radius relation. We lack a proper $N$-body simulation of the relative disruption rates for binary systems such as those that produce He WDs.

\subsubsection{Flaring}
Second, we consider the relative rate of luminous flares arising from the disruption of He WDs. One consideration is that He WD disruptions will produce a higher peak luminosity relative to MS stars, simply because they are more compact. For $0.5~M_\sun$ WDs, \citet{2014ApJ...794....9M} showed that their disruption rate $\dot N$ is lower than that of MS stars, but that, when weighted by their luminosities, the total number of observed transients is higher for these WDs than MS stars for $M_\text{bh} \lesssim 10^5~M_\sun$, as the observing volume grows with luminosity. For He WDs, one can similarly expect their luminosity-weighted rates to be higher relative to MS stars than their pure fractional rates estimated above. 

The fraction of prompt versus delayed flares is also important here. As suggested earlier, prompt flares occur when general relativistic effects are important. \citet{2015ApJ...809..166G} showed that MS stars are ineffectively circularized for lower BH masses, leading to viscously delayed luminosity evolution. For $10^5 < M_\text{bh}/M_\sun < 10^6$, the fraction of prompt events from MS stars is $\approx 13\%$ (if we include events that are viscously slowed only as they rise to peak, this fraction is $\approx 17\%$). Because He WDs are disrupted in the strongly relativistic regime, these objects should be rapidly circularized for these BH masses, as shown in Figure \ref{menu}. As a result, He WD disruptions should make up a higher fraction of prompt flares than their population fraction suggests. This effect becomes especially important at lower BH masses, for which the occupation fraction remains unconstrained \citep[e.g.][]{2010PhRvD..81j4014G}.

As we will see, even partial disruptions of He WDs with hydrogen envelopes can provide super-Eddington fallback onto the BH. These partial disruptions are also favorably prompt compared to MS disruptions, and could further enhance the relative rate of flares from He WD disruptions.

% ----------------------------------------------------------------------------
\section{NUMERICAL SETUP}\label{sec:setup}

\subsection{MESA Calculations}

Using the MESA stellar evolution code, we construct a $0.17~M_\sun$ white dwarf with a $0.16~M_\sun$ degenerate helium core and a $0.01~M_\sun$ hydrogen envelope. As noted in the previous section, there is a growing population of observed objects in this mass range. In these low-mass objects, the extended envelope lasts for a long time, as $\Delta t_\text{proto} \propto M_\text{WD}^{-7}$ (Equation \ref{eq:deltatproto}). We might therefore be more likely to see flares from the stripped envelopes of objects close to this mass, as they exist in a bloated state for longer than their higher-mass cousins.

\begin{figure}[tbp]
\epsscale{1.21}
\plotone{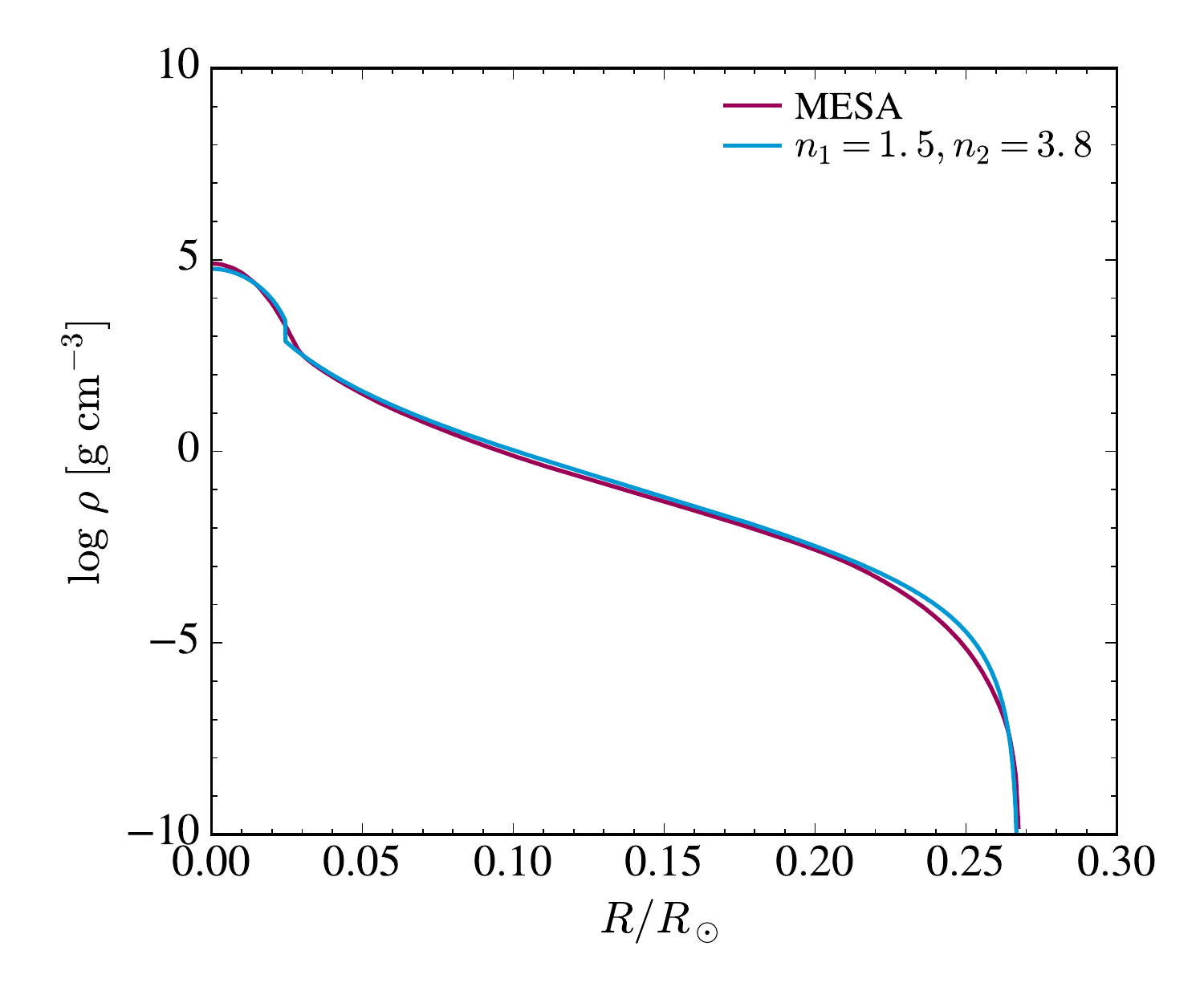}
\caption{Our matching of a nested polytrope with $n_\text{core}=1.5$ and $n_\text{env}=3.8$ to the MESA density versus radius profile of a $0.17~M_\sun$ WD with a $0.01~M_\sun$ hydrogen envelope. We use this nested polytrope in our tidal disruption calculations as our initial condition.}
\label{mesa}
\end{figure}

We approximate the core and envelope as nested polytropes \citep[e.g.,][]{1983ApJ...275..713R, 2012ApJ...757..134M, 2013ApJ...762...37L}, using polytropic indices $n_\text{core}=1.5$ and $n_\text{env}=3.8$. Figure \ref{mesa} shows the density versus radius profile of this object from MESA as well as from the nested polytrope that we matched. We use this nested polytrope as an input to our hydrodynamical simulations as it provides a reasonable description of the object's structure, and makes possible comparisons with non-hydrogen-envelope WD disruption calculations using polytropic equations of state \citep{2014ApJ...794....9M}.

A single polytrope is unstable to small variations in pressure $p_0$ and volume $V_0$ if $(\partial p/\partial V)_0$ is positive, and this occurs for polytropic indices of $n>3$. However, it is difficult to derive simple stability criteria for our nested polytrope structure, as it is not differentiable across the core-envelope discontinuity. We instead ensure that two heuristic tests of stability are satisfied: (1) the entropy increases with radius, or $\partial S/\partial r>0$, and (2) the star does not contract or relax significantly when placed on our hydrodynamical grid structure for 20 dynamical timescales of the full star. The dynamical timescale for the full star is $t_\text{dyn}^{\text{full}} \simeq \sqrt{R^3/GM} = 535$ s. In this work we will often refer to the dynamical timescale of the He core of this WD for comparison; this is $t_\text{dyn}^\text{core}=22.5$ s.

\subsection{Hydrodynamical Setup}

\begin{figure*}[tbp]
\epsscale{1.19}
\plotone{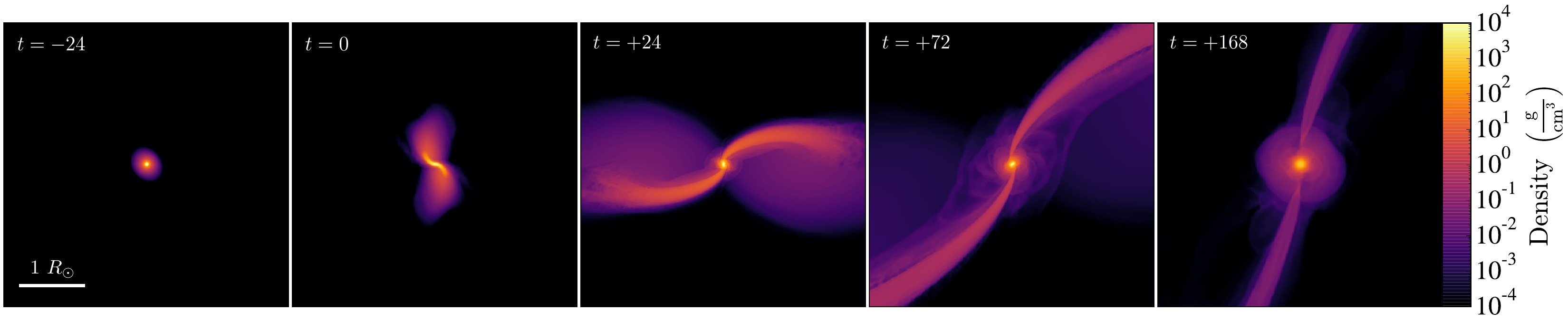}
\caption{
2D slices in density through our 3D simulation box, zoomed in on the star, for a $0.17~M_\sun$ He WD being disrupted by a $10^5~M_\sun$ BH. Panels from left to right show the time evolution for a $\beta_\text{core}=0.7$ encounter in units of the dynamical time of the core (22.5 s), with $t=0$ corresponding to pericenter.
}
\label{time_series}
\end{figure*}

\begin{figure*}[tbp]
\epsscale{1.1}
\plotone{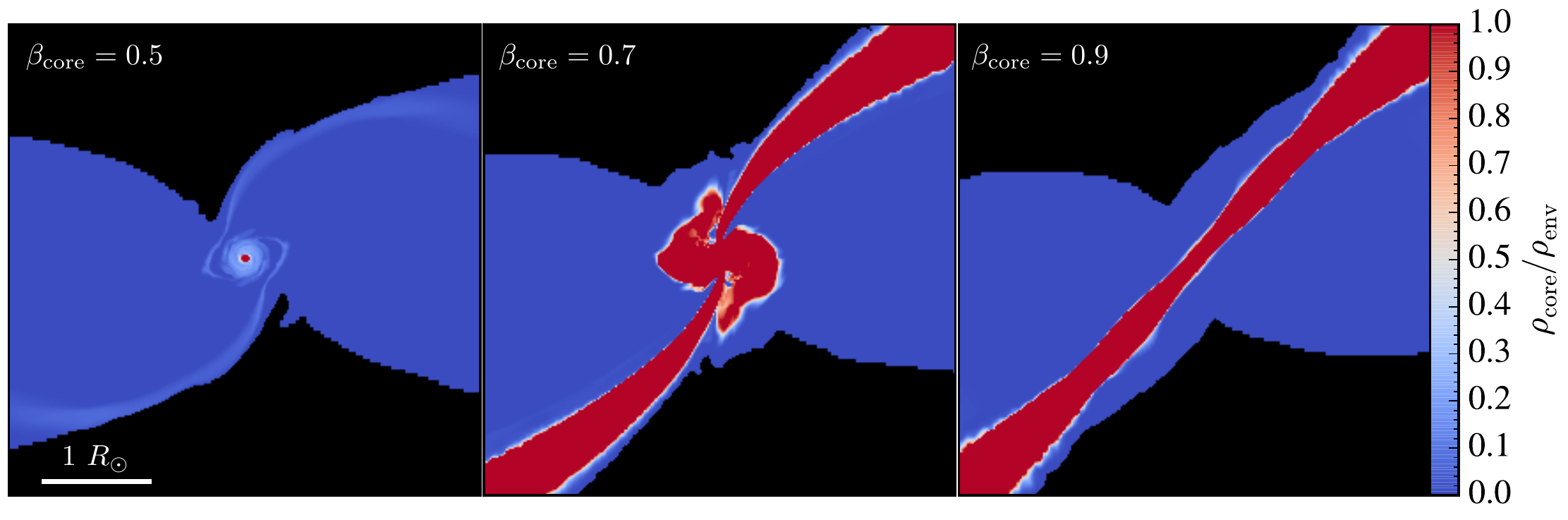}
\caption{
Panels from left to right show the mass fraction of core (red) versus envelope (blue) material for $\beta_\text{core}=0.5$, 0.7, and 0.9 encounters. These respectively correspond to a grazing encounter where just the envelope is stripped, an intermediate encounter, and full disruption. All slices are at $t=96\ t_\text{dyn}^\text{core}$ after pericenter. Density below $10^{-4}\ \mathrm{g/cm^3}$ is shown in black.
}
\label{beta_series}
\end{figure*}

Our simulations of tidal disruption are performed with the basic framework and code described in detail in \citet{2009ApJ...705..844G, 2011ApJ...732...74G}, \citet{2012ApJ...757..134M}, \citet{2013ApJ...762...37L}, and \citet{2013ApJ...767...25G}. We use FLASH \citep{2000ApJS..131..273F}, a 3D adaptive mesh grid-based hydrodynamics code including self-gravity. Hydrodynamics equations are solved using the using the piecewise parabolic method \citep{1984JCoPh..54..174C}. We refine the grid mesh on the value of the density, and derefine by one level every decade in density below $\rho=10^{-4}\ \mathrm{g\ cm^{-3}}$. All of the simulations presented here are resolved by at least $R_\star/\Delta r_\text{min}>130$, where $\Delta r_\text{min}$ is the size of the smallest cells. We note that adaptive mesh refinement is well suited for disruption calculations of an object with this core and envelope structure, as the envelope occupies a large volume yet has a very low mass fraction. 

We perform our calculations in the rest-frame of the star to avoid introducing artificial diffusivity by moving the star rapidly across the grid structure. We solve the self-gravity of the star using a multipole expansion about the center of mass of the star with $l_\text{max}=10$. We then evolve the orbit based on the center of mass of the star and the position of a point-mass black hole \citep[see the Appendix of][for details]{2011ApJ...732...74G}. We use Newtonian gravity for the black hole, which is a reasonable approximation as our star's closest approach in any of our simulations is $>10 r_\text{g}$, in the weak field regime. \citet{2014PhRvD..90f4020C} showed that general relativistic effects in tidal disruption simulations should be small is this regime. Note that, because we use Newtonian gravity, by construction, the encounters we simulate are outside of our rapid circulation condition defined in Section \ref{sec:menu}. The effect of relativistic encounters is discussed in Section \ref{subsec:caveats}.

We run our simulations using the $0.17~M_\sun$ He WD described above and a $10^5~M_\sun$ BH. We input the MESA profile, matched as a nested polytrope (Figure \ref{mesa}), into FLASH. We use two different fluids in the simulation: one for the helium core and one for the hydrogen envelope. Both have the same equation of state, with a $\gamma_\text{fluid}=5/3$. This setup has an envelope composition of 100\% hydrogen. More accurately, the envelope has a residual helium abundance that will migrate toward the core over time depending on the relative strength of mixing and gravitational settling. We relax the object onto the grid for $5\ t_\text{dyn}$ before sending the BH toward it. We use an eccentricity $e\approx1$, as most disrupted stars originate from orbits scattered from the sphere of influence \citep{1999MNRAS.309..447M, 2004ApJ...600..149W}. As discussed in \citet{2013ApJ...767...25G}, for a given stellar structure, we can understand the vast majority of disruptions by surveying in impact parameter $\beta=r_\text{t}/r_\text{p}$ as all other parameters obey simple scaling relations when relativistic effects are unimportant. Similar to the dynamical timescale, we can define $\beta$ with respect to the tidal radius of the full star or the degenerate core. We survey in $\beta_\text{full}$ from 1 to 10 in 12 runs. This corresponds to $\beta_\text{core}$ of $\approx 0.1$ to 1.2. We run our simulations for $21\ t_\text{dyn}^\text{full}=500\ t_\text{dyn}^\text{core}$, well into the self-similar decay portion of the mass fallback rate.

% ---------------------------------------------------------------------------
\section{NUMERICAL RESULTS}\label{sec:results}

\subsection{Phenomenology: Core versus envelope}

Figure \ref{time_series} shows the time evolution of the star for a $\beta_\text{core}=0.7$ encounter in 2D slices in density through the 3D simulation box, zoomed in on the star. Time is labeled in terms of the dynamical time of the core. In this moderately plunging encounter, the star is distorted through pericenter, evolving into a surviving remnant and two tidal tails---one bound and one unbound from the BH.

As we increase the impact parameter, the star is perturbed closer to its center. For mildly plunging encounters, only the hydrogen envelope is stripped, while the core survives intact. For more deeply plunging encounters, both the core and envelope are disrupted and fed to the BH. We can see this qualitatively in Figure \ref{beta_series}, where we show slices through the simulation box zoomed in on the star for $\beta_\text{core}=0.5$, 0.7, and 0.9 encounters. We plot the ratio of the core material to envelope material density. The different spatial distributions of core and envelope material will result in different fallback times to the BH, which will result in observed light curves dominated by material of different compositions at different times. Because of their different structures---the envelope has a steeper density gradient than the core---these two fluids react to losing mass in characteristically different ways, as we will see below.

\subsection{Mass lost}

Figure \ref{deltaM_vs_beta_poly} shows the mass lost from the star as a function of impact parameter, calculated at the last timestep of our simulations. We run our simulations long enough so that the mass lost calculated from this final timestep is asymptotically close to the final mass lost. Note that half of the lost mass will return to the black hole and half is ejected as an unbound debris stream. The object is smoothly disrupted with the impact parameter, albeit with two components from the envelope and the core. This is different from giant star disruptions \citep{2012ApJ...757..134M}, where the core is never disturbed, and likely arises because the density contrast between the core and envelope is in general larger for giants than it is for He WDs.

A fitting formula from \citet{2013ApJ...767...25G, 2015ApJ...798...64G} for a $\Gamma=5/3$ polytrope fits the mass lost from the core well. This is expected, as once the core has been penetrated, the envelope has negligible dynamical effect, and the disruption will proceed as if for a typical WD. Full disruption occurs at $\beta_\text{core} \approx 0.9$.

This $n=1.5$ polytrope has a lower critical $\beta$ (for full disruption) compared to higher index polytropes, as the mass is distributed more evenly. In addition to this, a $n=1.5$ polytrope has an inverse mass-radius relation, and so expands when mass is removed, making the object more vulnerable to disruption. We model the envelope, on the other hand, as an $n=3.8$ polytrope, which reacts to mass removal by contracting---``protecting'' itself. Because the envelope has a steeper density gradient, its critical $\beta$ is higher than for a $\Gamma=5/3$ polytrope. We can see this in the shallower slope of $\Delta M/M$ versus $\beta$ for envelope material relative to core material.

\begin{figure}[tbp]
\epsscale{1.24}
\plotone{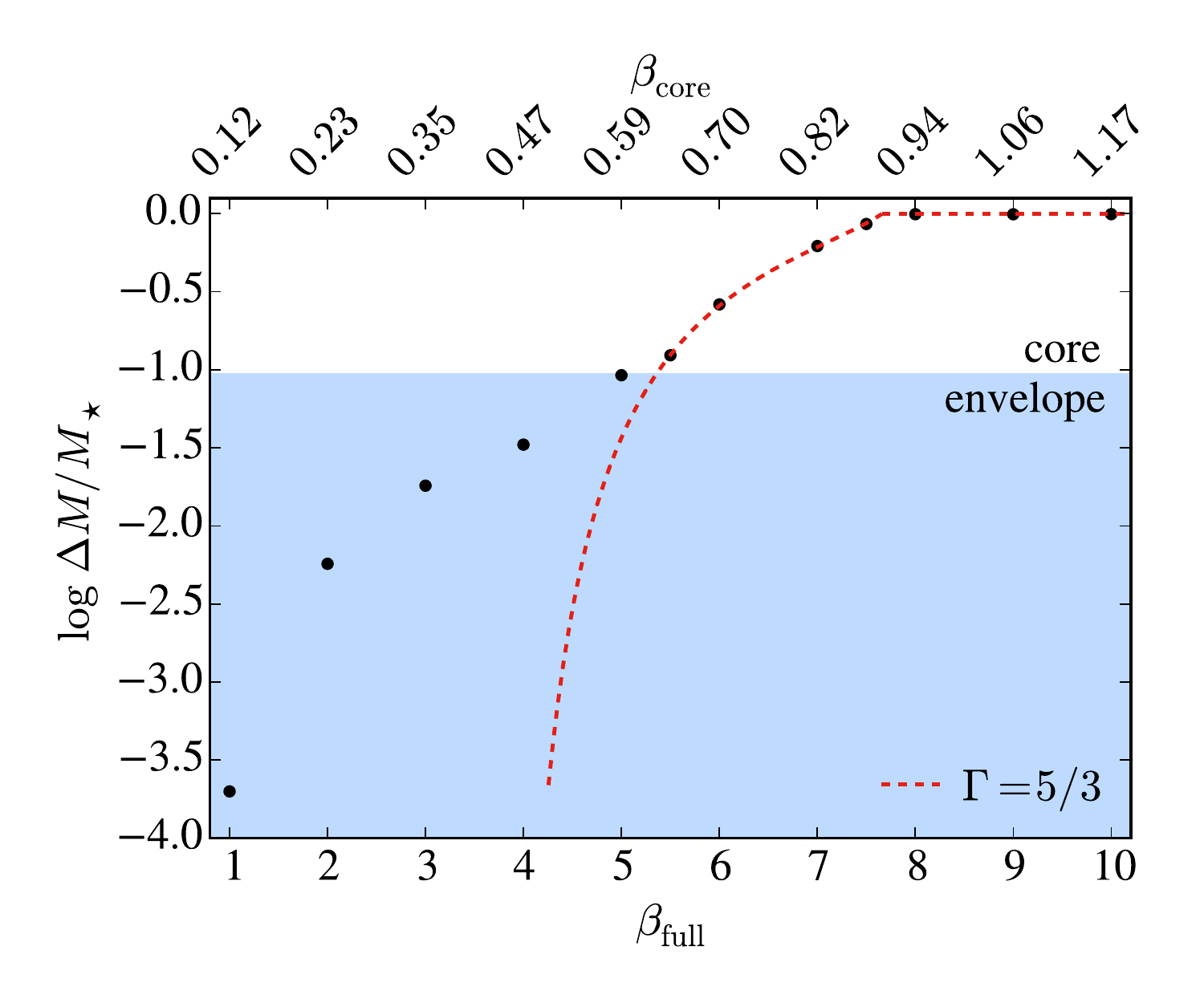}
\caption{
Mass lost versus impact parameter for the disruption of a $0.17~M_\sun$ He WD with a $10^5~M_\sun$ BH is shown in solid circles. The total mass of the envelope is shaded in blue. A fitting formula from \citet{2013ApJ...767...25G} for a $\Gamma=5/3$ polytrope is shown in dashed red. Once the core begins to be disrupted as well, the single 5/3 polytrope yields the same amount of mass loss as the nested polytrope. The bottom $x$-axis shows the $\beta$ of the full star (i.e., including the envelope), and the top $x$-axis shows the equivalent $\beta$ of only the core.
}
\label{deltaM_vs_beta_poly}
\end{figure}

\subsection{Spread in binding energy and mass fallback rate}

We calculate the spread in binding energy of the star's material to the BH, $dM/dE$ versus $E$, over time. We compute the specific binding energy of the material in each cell of the simulation, which depends on its distance and velocity relative to the center of mass of the star and to the black hole. Details of the calculation are presented in \citet{2013ApJ...767...25G}. Only material that is bound to the BH and not bound to the star will contribute to the mass fallback onto the BH. We compute the specific binding energy of the material in each cell of the simulation, which depends on its distance and velocity relative to the center of mass of the star and to the black hole.

\begin{figure*}[tbp]
\epsscale{0.74}
\plotone{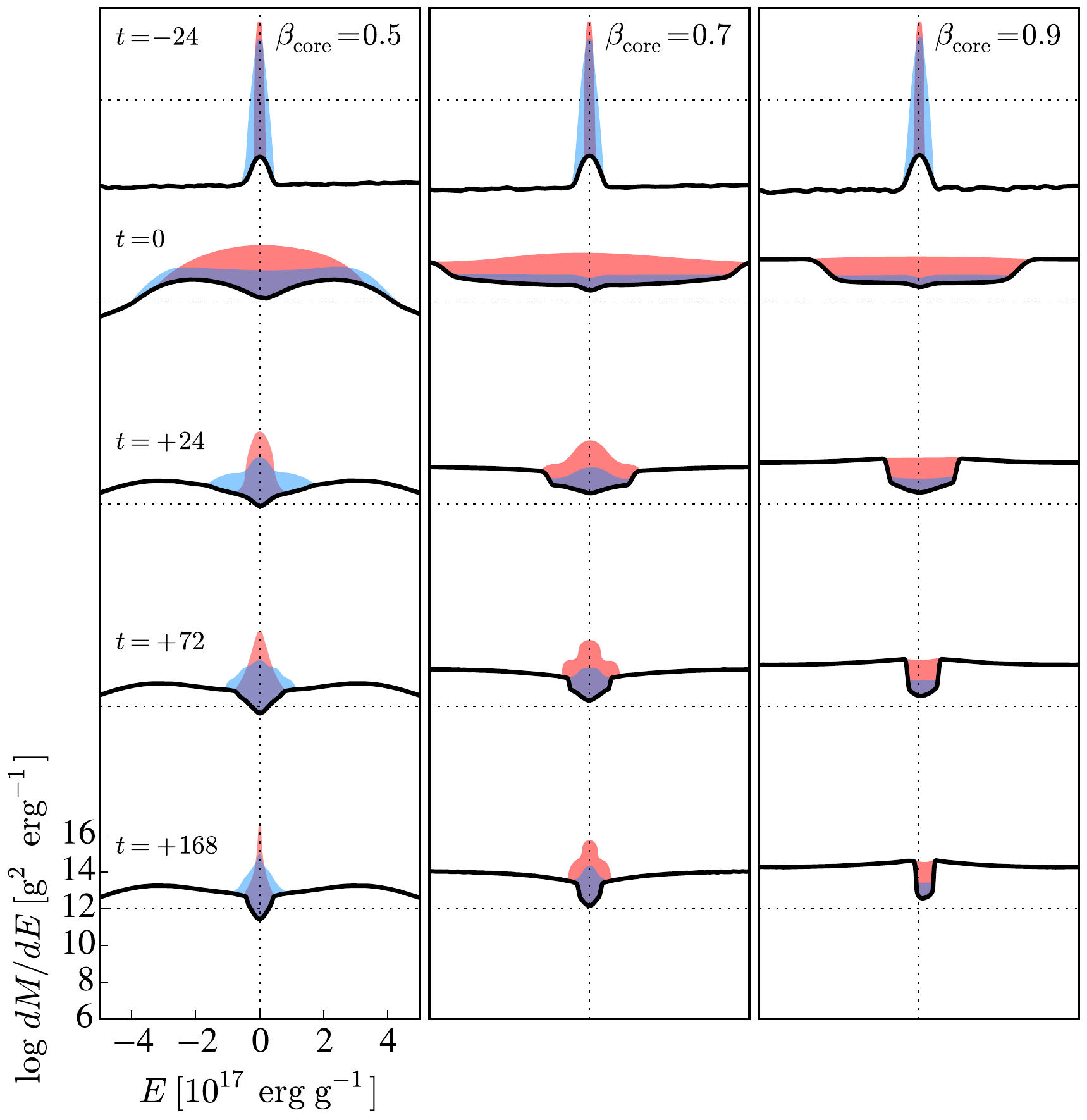}
\caption{
Panels from left to right show the spread in specific binding energy for $\beta_\text{core}=0.5$, 0.7, and 0.9 encounters (the same as shown in Figure \ref{beta_series}). Time increases from top to bottom for each impact parameter and is labeled in terms of $t_\text{dyn}^\text{core}$. Material bound to the star (the remnant) is shown in red and blue, corresponding to core and envelope material, respectively. Material unbound to the star (the tidal tails) is shown by the black solid line. A vertical dashed line is shown for reference at $E=0$, where material is moving with the center of mass of the star. A horizontal dashed line is shown for reference at $dM/dE=10^{12}~{\rm g^2~erg^{-1}}$. The binding energy of material both bound and unbound to the star varies widely with $\beta$. Material spreads out in binding energy through disruption, and higher impact parameters spread out the binding energy more effectively. Higher impact parameters also leave a more compact remnant. A grazing encounter retains some of the envelope, while a deeply plunging encounter loses nearly all of it.
}
\label{pink}
\end{figure*}

Figure \ref{pink} shows the spread in $dM/dE$ over time for $\beta_\text{core}=0.5$, 0.7, and 0.9 encounters, with the contribution from material unbound to the star in solid black and contributions from the core and envelope of the remnant in red and blue. We see that impact parameter drastically changes the spread in binding energy through and following disruption, both for the bound and unbound material. Grazing encounters leave the core relatively unperturbed and are able to retain more envelope material, while deeper encounters leave a compact remnant that has been all but stripped of its envelope.

\begin{figure*}[tbp]
\epsscale{1.0}
\plottwo{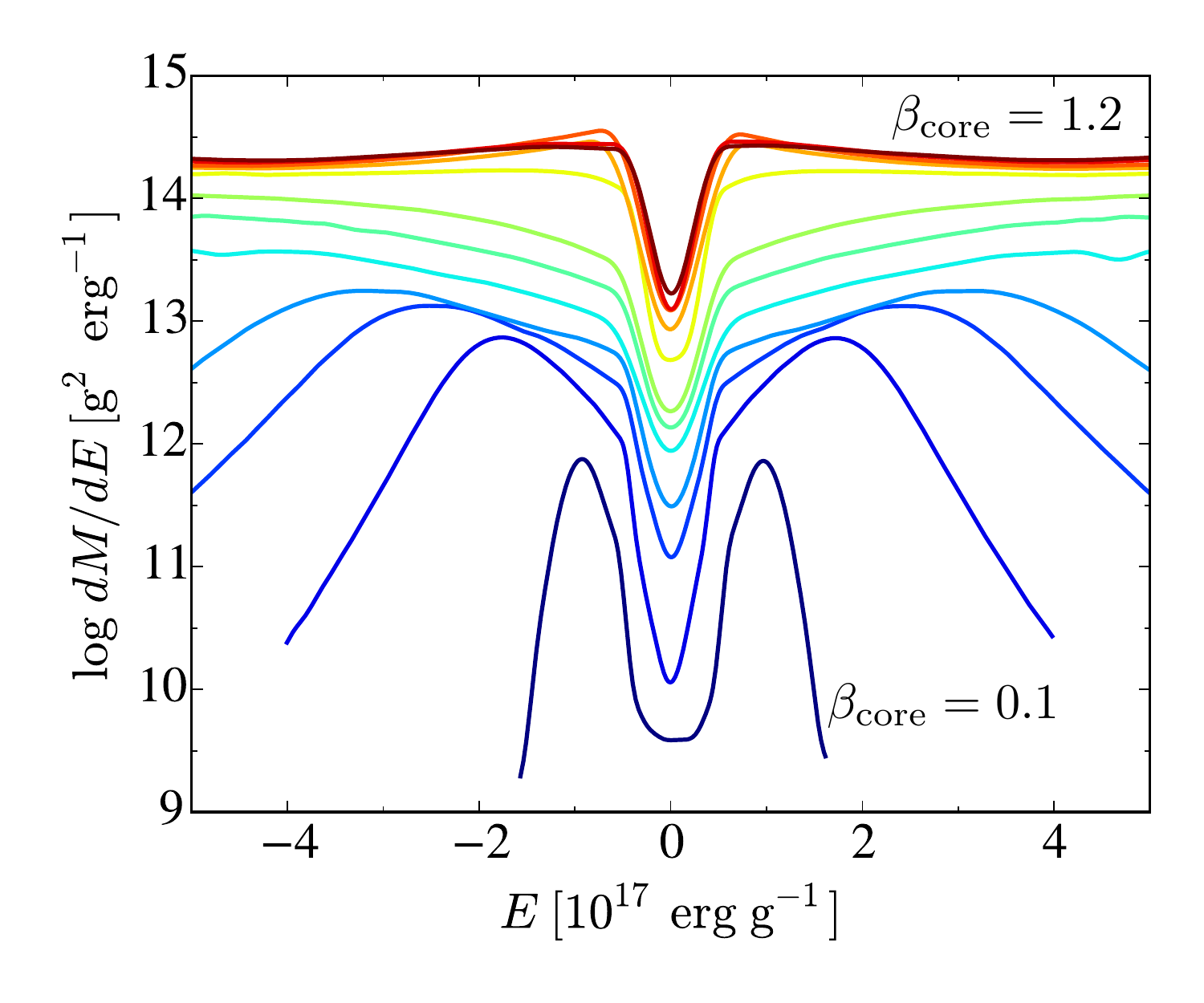}{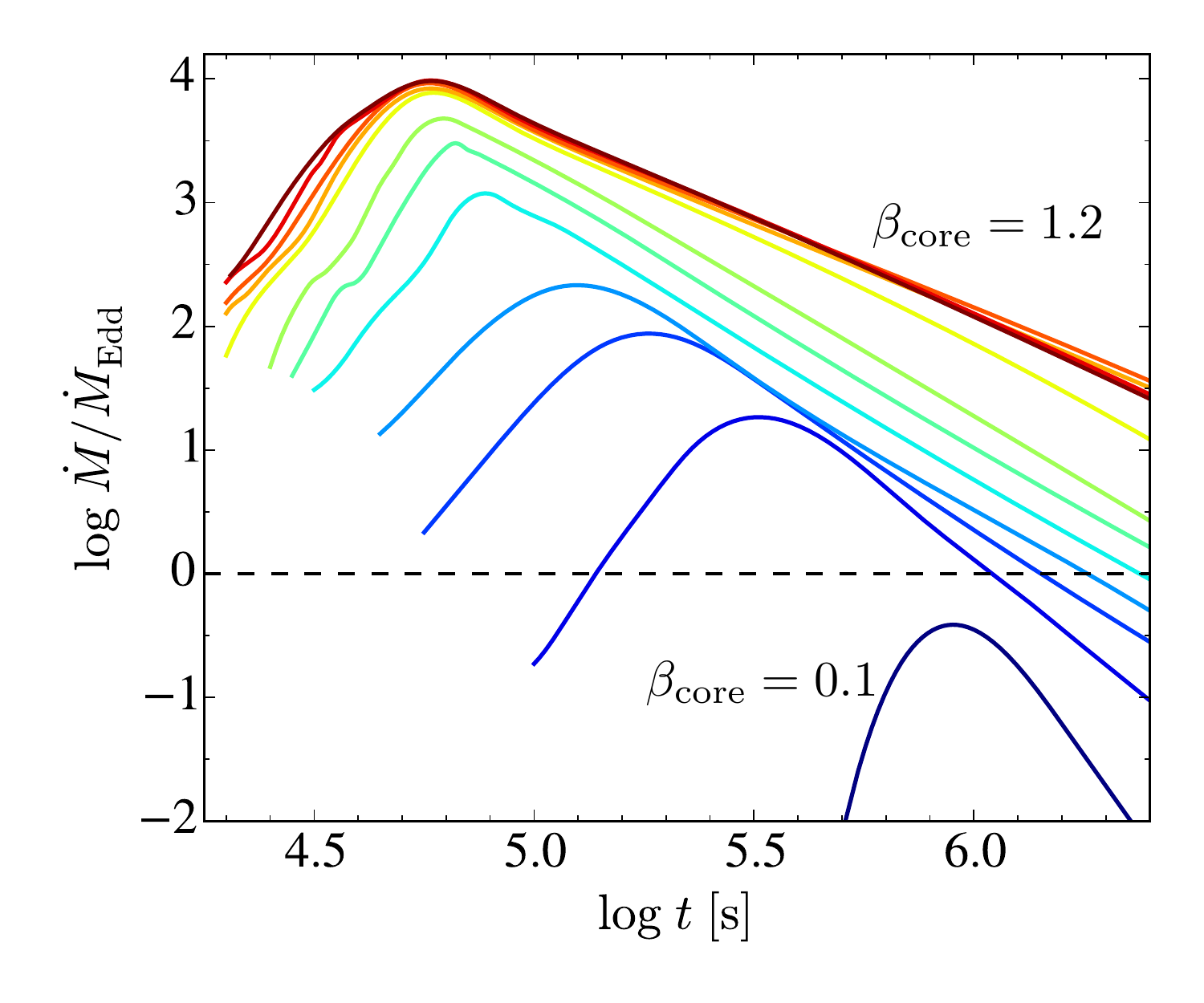}
\caption{
Left panel: spread in specific binding energy as a function of $\beta$ for a $0.17~M_\sun$ He WD disrupted by a $10^5~M_\sun$ BH. Right panel: mass fallback rate $\dot M$ onto the BH versus time for the same impact parameters, with the Eddington limit for this BH shown in dashed black. Impact parameters range from $\beta_\text{core}=0.1$ to 1.2 in increments of roughly 0.1. See Figure \ref{deltaM_vs_beta_poly} for the corresponding mass lost for each $\beta$.
}
\label{mdot}
\end{figure*}

Given $dM/dE$ and a pericenter distance, we can calculate the mass fallback rate onto the BH by Kepler's third law,
\begin{equation}
\frac{dM}{dt} = \frac{dM}{dE} \frac{dE}{dt} = \left(\frac{dM}{dE}\right) \frac{1}{3} \left(2\pi GM_\text{bh}\right)^{2/3} t^{-5/3}.
\end{equation}
The left panel of Figure \ref{mdot} shows the spread in specific binding energy $dM/dE$ versus $E$ at the last timesteps of our simulations for all impact parameters. We verify that the binding energy has effectively ``frozen in,'' or converged to its final distribution, by this timestep. The right panel shows $dM/dE$ mapped onto $dM/dt$ across time for the same impact parameters, with the Eddington limit for this BH shown in dashed black. We take $\dot M_\text{Edd}=0.02\ (\eta/0.1)\ (M_\text{bh}/10^6~M_\sun)\ \mathrm{M_\sun/yr}$ with $\eta=0.1$. Feeding rates peak at $t_\text{peak} \sim 5 \times 10^4\ \text{to}\ 10^6\ \text{s} \approx 0.6\ \text{to}\ 11\ \text{d}$ depending on $\beta$. Weakly plunging encounters peak later, while deeply plunging encounter peak earlier. Note that $t_\text{peak}$ evolves strongly with $\beta$ (it spans more than an order of magnitude), in contrast to single polytrope solutions where the evolution in $t_\text{peak}$ is much more gradual \citep[e.g.,][]{2013ApJ...767...25G}. This means that the He WD disruptions---and disruptions of other objects with this core and extended envelope structure---probe a much wider range of potential transient characteristics for a given BH mass. We see that even for very weakly plunging encounters, for which only a fraction of the envelope is stripped (see mass lost in Figure \ref{deltaM_vs_beta_poly}), the mass fallback rate is super-Eddington. Encounters only stripping the envelope appear to have a shallower slope in early-time mass fallback and smoother evolution near peak than encounters penetrating the core. This is due to their different polytropic structures.

\begin{figure}[tbp]
\epsscale{1.21}
\plotone{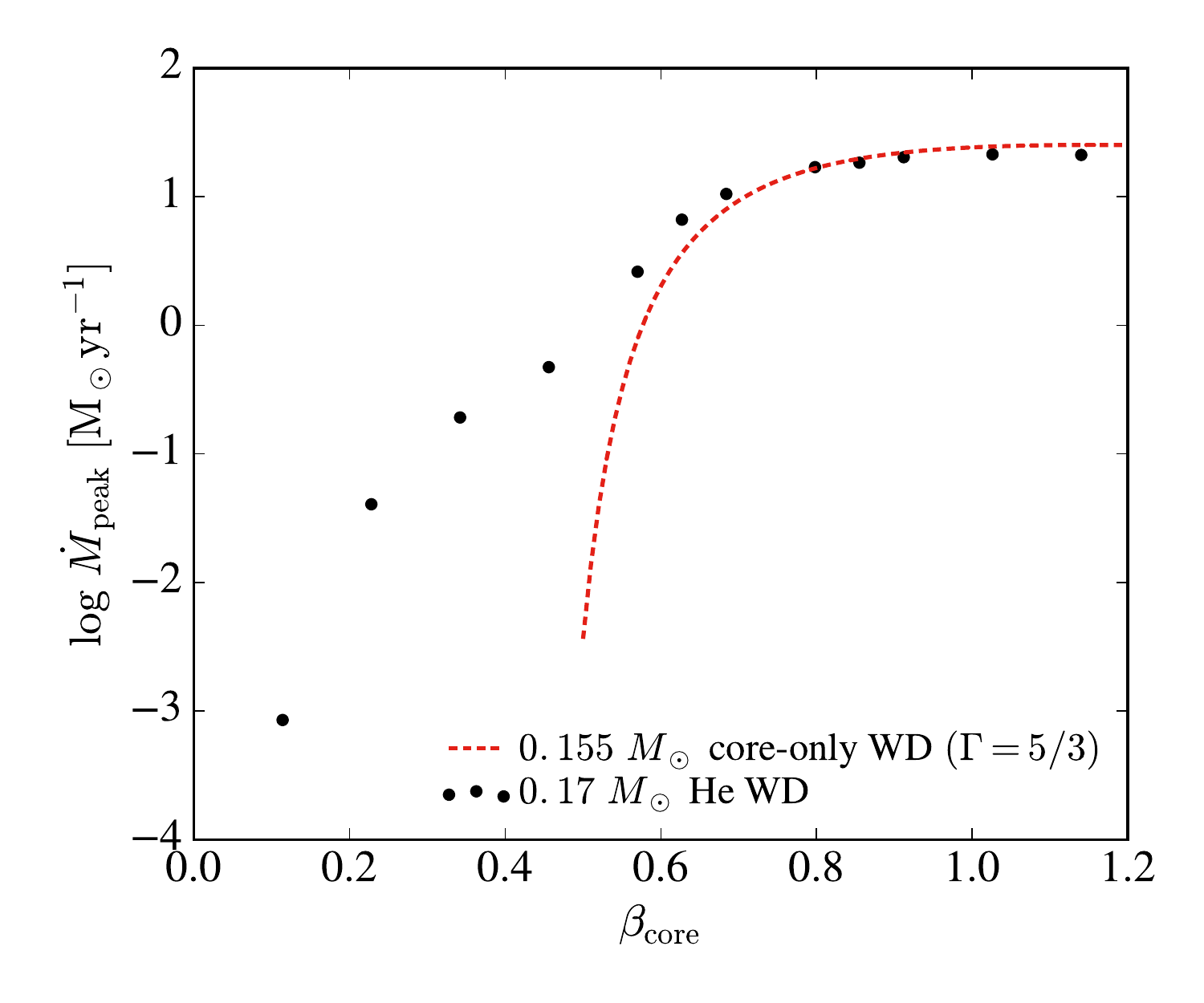}
\caption{
Black filled circles show the peak mass fallback rate, $\dot M_\text{peak}$, versus $\beta$ for the encounters shown in Figure \ref{mdot}. The $\dot M_\text{peak}$ fitting formula from \citet{2013ApJ...767...25G} for a $\Gamma=5/3$ polytrope with the mass of the core of this object, $0.155~M_\sun$, is shown in dashed red. $\dot M_\text{peak}$ values for encounters that penetrate the core are close to those of a non-hydrogen-envelope WD, while low $\beta$ encounters provide fallback rates unavailable to WDs without envelopes.
}
\label{mdotpeak_vs_beta_compare53}
\end{figure}

\citet{2013ApJ...767...25G} presented a fitting formula for the peak fallback rate of material onto the BH, where $\dot M_\text{peak} = f(M_\text{bh}, \beta, \gamma)$. Figure \ref{mdotpeak_vs_beta_compare53} shows $\dot M_\text{peak}$ values from our simulations of the disruption of an $0.17~M_\sun$ He WD compared with those from this fitting formula for a $\Gamma=5/3$ non-hydrogen-envelope WD with a mass of $0.155~M_\sun$, the mass of the core of the He WD. We expect this functional form to match for disruptions that penetrate the core. In low $\beta$ encounters, the hydrogen envelope provides mass return rates that are unavailable to WDs without envelopes.

\subsection{Composition of debris}

\begin{figure*}[tbp]
\epsscale{1.18}
\plotone{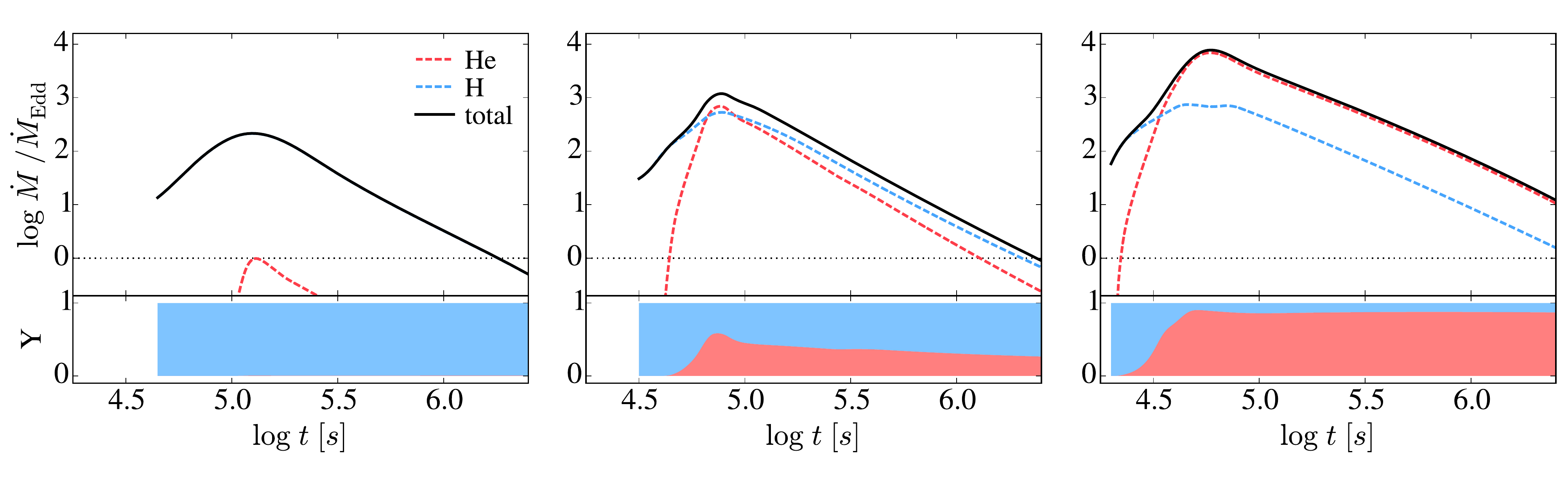}
\caption{
The tops of the panels from left to right show the mass fallback rate as a function of time for $\beta_\text{core}=0.5$, 0.6, and 0.8 encounters, respectively. The total $\dot M$ is shown in solid black, and the Eddington limit for this BH in dotted black. Contributions from the helium core and hydrogen envelope are shown in dashed red and blue, respectively. The bottoms of the panels show the mass fraction of $\dot M$ over time from helium and hydrogen. The mass fallback rate from weakly plunging encounters can be super-Eddington and hydrogen-dominated. In more deeply plunging encounters, there is a transition between envelope-fed fallback and core-fed fallback that depends on $\beta$.
}
\label{composition}
\end{figure*}

We track the core and envelope material separately in our simulations, which allows us to track the composition of the debris falling onto the BH. In Figure \ref{composition} we show $\dot M$ as a function of time for $\beta_\text{core}=0.5$, 0.6, and 0.8, with absolute and fractional contributions from the helium core in red and the hydrogen envelope in blue. The mass fallback rate from weakly plunging encounters can be super-Eddington and hydrogen-dominated. In more deeply plunging encounters, the early rise of the mass fallback rate is fed almost entirely by the hydrogen envelope, while the peak and late time evolution are fed by the helium core; the nature of this transition depends on $\beta$. Note that the disruption turns the star inside out: the material that is removed first accretes first, and is then buried underneath the material that is removed last and accretes last. The diffuse envelope material feeds a qualitatively slower rise in the mass fallback curve compared to the core material.

The first evidence that a range of stellar or spectral properties might be represented in TDEs was the discovery of a helium-rich TDE, PS1-10jh \citep{2012Natur.485..217G}. Gezari et al. explained its hydrogen-free spectrum as the result of the tidal disruption of the helium-rich core of a star, similar in structure to an He WD progenitor. \citet{2014ApJ...793...38A} noted that TDEs observed thus far show a continuum of helium-rich to hydrogen-rich spectral features; there is an ongoing debate over the origin of the strong helium emission. \citet{2016MNRAS.458..127K} found that stellar evolution can play a role in producing this spectral diversity. \citet{2016ApJ...827....3R} modeled the emission from TDEs through an extended, optically thick envelope formed from stellar debris. They find that due to optical depth effects, hydrogen Balmer line emission is often strongly suppressed relative to helium line emission. For MS stars, for example, it is possible for the hydrogen emission lines to be absent. Having said this, the specific composition of the material is expected to have consequences on the detailed line ratios. Line diagnostics from disruptions of He WDs could transition from hydrogen to helium smoothly with $\beta$. An encounter stripping only the envelope could provide a rare, (nearly) pure hydrogen-powered mass fallback. If an optically thick reprocessing envelope exists, however, observational evidence of this type of encounter could be variable.

% ----------------------------------------------------------------------------
\section{TDE DEMOGRAPHICS}\label{sec:demographics}

Here we explore the tidal disruption menu of BHs and disrupted objects in terms of the peak fallback rate and its associated peak timescale, and place our He WDs in context. Through Kepler's third law, we can write scalings of the peak mass fallback rate and its associated time of peak,
\begin{align}
\dot M_\text{peak} &\propto M_\text{bh}^{-1/2}\ M_\star^{2}\ R_\star^{-3/2} \\
t_\text{peak} &\propto M_\text{bh}^{1/2}\ M_\star^{-1}\ R_\star^{3/2},
\end{align}
where the $\dot M_\text{peak} \propto M_\star^{2}$ scaling results when we assume that a constant fraction of the star's mass is lost in the disruption. \citet{2013ApJ...767...25G, 2015ApJ...798...64G} found fitting parameters for these scaling relations that depend on the polytropic $\Gamma$ and impact parameter $\beta$. We use these below.

\begin{figure}[tbp]
\epsscale{1.23}
\plotone{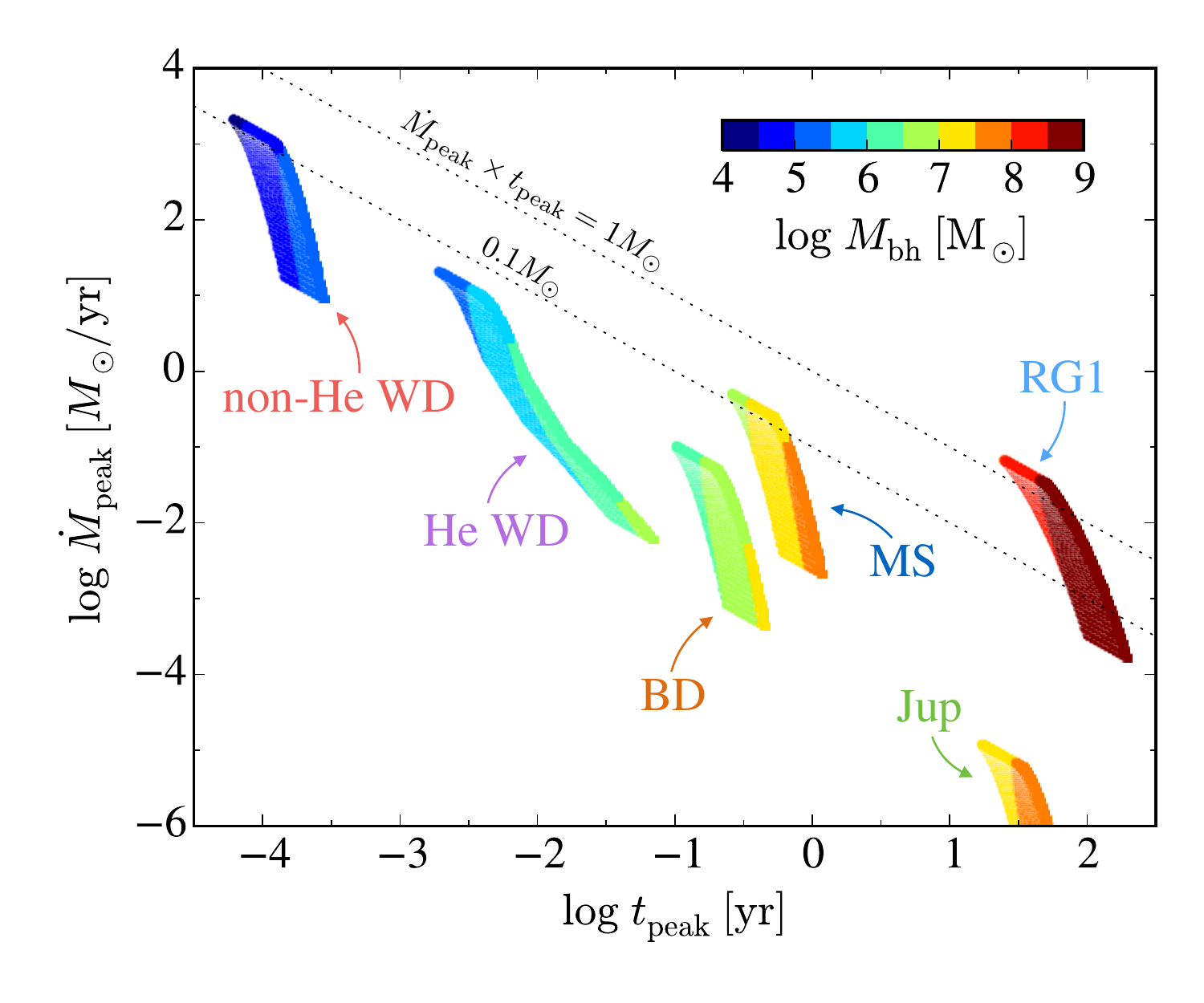}
\caption{
Peak mass fallback rate versus time of peak for a $0.6~M_\sun$ non-He WD, a $0.17~M_\sun$ He WD, a $0.6~M_\sun$ MS star, a $50~M_\text{Jup}$ brown dwarf, a $1~M_\text{Jup}$ planet, and a $1.4~M_\sun$ red giant at RG1 ($\approx 10~R_\sun$). Encounters are colored by BH mass. Dotted lines show where $\dot M_\text{peak} \times t_\text{peak} = 0.1~M_\sun$ and $1~M_\sun$. We show only encounters obeying our circularization requirement, $4GM_\text{bh}/c^2 < r_\text{t} < 10 GM_\text{bh}/c^2$, favoring prompt flares. We show impact parameters that remove from $\Delta M/M_\star=0.01$ to 1 from each object.
}
\label{mdotpeak_vs_tpeak_scatter}
\end{figure}

In Figure \ref{mdotpeak_vs_tpeak_scatter}, we show $\dot M_\text{peak}$ versus $t_\text{peak}$ values for the He WD disruptions presented in this work, as well as for several representative disruptions of other objects: a $0.6~M_\sun$ non-He WD, a $0.6~M_\sun$ MS star, a $50~M_\text{Jup}$ brown dwarf (BD), a $1~M_\text{Jup}$ planet, and a $1.4~M_\sun$, $10~R_\sun$ red giant (RG). We use fitting parameters from Equations A1 and A2 of \citet{2013ApJ...767...25G} to calculate $\dot M_\text{peak}$ and $t_\text{peak}$ for the other objects, and to scale with BH mass. We use a polytropic $\Gamma$ of $5/3$ for the WD, MS star, BD, and planet (the values are similar if we use $4/3$ for the MS star), and $4/3$ for the RG. We show impact parameters that remove from $\Delta M/M_\star=0.01$ to 1 from each object.

As in the tidal disruption menu shown in Figure \ref{menu}, we only show encounters with BHs obeying our prompt circularization condition, $4GM_\text{bh}/c^2 < r_\text{t} < 10 GM_\text{bh}/c^2$. Here, flares resulting from the fallback of material onto the BH are both visible (disruption occurs outside the innermost bound circular orbit) and predominantly prompt (circularization of the debris is efficient). We color the encounters by BH mass. 

There is a huge variety in the timescales and fallback rates with which stars feed MBHs following TDEs. Prompt flares separate into different timescale classes based on the stellar type and BH mass combination. Prompt flares also show relatively unique timescale/BH mass combinations---i.e., a timescale and a prompt flare can imply not only a stellar type but also a BH mass. The clean separations blur slightly if we allow for (1) the full distribution of masses and radii available for different classes of objects, which is especially important for He WDs, and (2) the effects of viscous delay, which smear the effective timescales and mass fallback rates.

\begin{figure}[tbp]
\epsscale{1.21}
\plotone{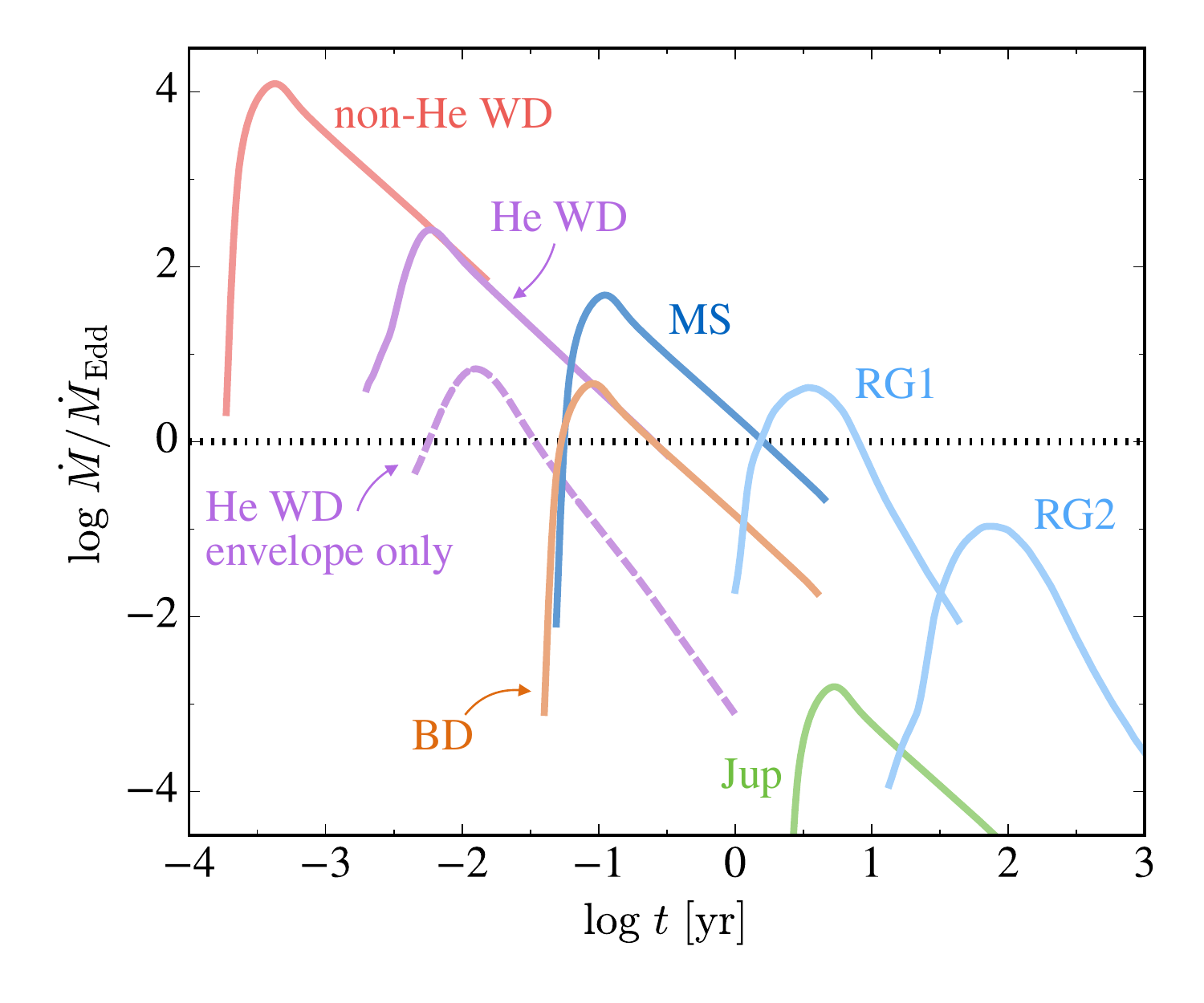}
\caption{
Mass fallback rate curves for the representative objects shown in Figure \ref{mdotpeak_vs_tpeak_scatter} scaled to a single BH mass ($10^6~M_\sun$) for comparison. Colors are the same as in Figure \ref{menu} menu. We show a $0.6~M_\sun$ non-He WD in red, a $0.17~M_\sun$ He WD in purple, a $0.6~M_\sun$ MS star in blue, a $50~M_\text{Jup}$ brown dwarf in brown, a $1~M_\text{Jup}$ planet in green, and a $1.4~M_\sun$ red giant at RG1 ($\approx 10~R_\sun$) and at RG2 ($\approx 100~R_\sun$) in light blue. We show a $\beta=0.9$ encounter (full disruption) for the non-He WD, MS star, BD, and planet, and a $\beta=1.5$ encounter for the giant stars. For the He WD, we show two $\dot M$ curves for comparison: the solid line shows a full disruption ($\beta_\text{core}=0.9$) and the dashed line shows an envelope-stripping encounter ($\beta_\text{core}=0.5$).
}
\label{mdots_menu}
\end{figure}

In Figure \ref{mdots_menu} we show a fallback rate curve for each of the objects in Figure \ref{mdotpeak_vs_tpeak_scatter}, scaled to disruptions with a $10^6~M_\sun$ BH for comparison. Note that the $0.6~M_\sun$ WD disruption would occur inside the event horizon for this BH mass. We show fallback from a $1.4~M_\sun$ RG at two different points along the giant branch: ascending the RG branch (RG1; $R \approx 10~R_\sun$) and the tip of RG branch (RG2; $R \approx 100~R_\sun$), from \citet{2012ApJ...757..134M}. We show a $\beta=0.9$ encounter (full disruption) for the non-He WD, the MS star, and the planet, and a $\beta=1.5$ encounter for the giant stars. We show two $\dot M$ curves for the He WD: one for a full disruption ($\beta_\text{core}=0.9$) and one for an envelope-stripping encounter ($\beta_\text{core}=0.5$). For a given BH mass, these objects offer distinct fallback rates and characteristic timescales. 

Converting these fallback rates into luminosities is not straightforward. In this paper we have focused on rapidly circularized TDEs, where the accretion rate (and so the luminosity) is expected to closely follow the fallback rate. This is predicted to be true for emission both from the disk \citep{2014ApJ...783...23G} and from stream collisions \citep{2015ApJ...812L..39D}, and is observed to be the case in the best-sampled, non-beamed UV/optical events \citep[e.g.,][]{2012Natur.485..217G, 2014ApJ...783...23G}. However, it is not evident that the luminosity will always follow the fallback rate, in particular when circularization is inefficient or for BHs accreting at highly super-Eddington rates \citep{2015ApJ...804...85S, 2015ApJ...806..164P}. For example, the event Sw J1644+57 \citep{2011Sci...333..203B} did not appear to follow a $t^{-5/3}$ luminosity evolution during its prompt decline phase. In addition, jetted emission may not be Eddington limited; its strength depends on the radiative efficiency of the (relativistic) flow. Even in the absence of a jet, \citet{2015MNRAS.454L...6M} show that the radiative efficiency of super-Eddington accretion flows can be high under certain circumstances.

While most full disruptions are expected to provide super-Eddington accretion rates (Figure~\ref{mdots_menu}), the observed peak luminosities of UV/optical TDEs appear to be Eddington limited, or sub-Eddington \citep{2017arXiv170301299H}. Two possible solutions to this are (1) that the most commonly observed events are partial disruptions, where the fallback rate can be significantly lower \citep[e.g.,][]{2013ApJ...767...25G} or (2) that the radiative efficiency is low \citep[e.g.,][]{2015MNRAS.453..157P}.

Constructing this menu---which spans many orders of magnitude in BH mass, fallback timescale, and fallback rate---is nonetheless a key step toward making meaningful comparisons with observations. We have only shown a few representative objects; the full phase space of luminosities and timescales, the effects of viscous delay, and a comparison to observations will be explored in future work.

% ----------------------------------------------------------------------------
\section{DISCUSSION}\label{sec:discussion}

\subsection{Possible Candidates for He WD Disruption}

Here we compare $t_\text{peak}$ values from simulations to those of two particularly rapidly rising TDE candidates, Dougie \citep{2015ApJ...798...12V} and PTF10iya \citep{2012MNRAS.420.2684C}, accounting for luminosity and BH mass constraints. \citet{2015ApJ...798...12V} estimate Dougie's peak bolometric luminosity as $L_\text{peak} \approx 5(\pm1) \times 10^{44}~\mathrm{erg~s^{-1}}$ and its rise time as $t_\text{rise}\sim10~\text{d}$. They estimate a central BH mass of a few $10^6$ to $10^7~M_\sun$ for Dougie's host galaxy. \citet{2012MNRAS.420.2684C} estimated 10iya's peak bolometric luminosity as $L_\text{peak} \approx (1-5) \times 10^{44}~\mathrm{erg~s^{-1}}$ and place a limit on its rise time of $t_\text{rise}<5~\text{d}$. They constrain the central BH mass via the observed bulge luminosity versus BH mass relation as $\log M_\text{BH}/M_\sun \lesssim 7.5$.

In order to constrain the kinds of disruptions that can produce such rapid flares, we construct a histogram of $t_\text{peak}$ for the $0.17~M_\sun$ He WD disruptions presented in this work as well as for regular WDs, MS stars, BDs, and planets. We model regular WDs, MS stars, BDs, and planets with $M < 0.3~M_\sun$ as $\Gamma=5/3$ polytropes. We model MS stars with $M > 0.3~M_\sun$ as $\Gamma=4/3$ polytropes. The mass at which we transition from $5/3$ to $4/3$ does not affect our conclusions significantly, as their $t_\text{peak}$ values overlap. Giant star disruptions have longer timescales than we are interested in here.

We draw from flat distributions in $M_\text{obj}$, with white dwarf masses of $0.2~M_\sun < M_\text{WD} < 1~M_\sun$, MS star masses of $0.085~M_\sun < M_\text{MS} < 3~M_\sun$, BD masses of $13~M_\text{Jup} < M_\text{BD} < 0.085~M_\sun$, and planet masses of $1~M_\text{Jup} < M_\text{pl} < 13~M_\text{Jup}$. We use only the one $0.17~M_\sun$ He WD mass. We draw from a flat distribution in BH mass with $10^6 < M_\text{bh}/M_\sun < 10^7$, roughly the BH mass constraints for Dougie and 10iya. We draw from a flat distribution in $\beta$, discarding encounters where $r_\text{p}<r_\text{ibco}$. We estimate the peak luminosity from each encounter as $L_\text{peak} = \text{min}\left(0.1 \dot M_\text{peak} c^2,\ L_\text{Edd}\right)$ for the given BH, as these events were observed in the optical/UV and we expect accretion luminosity to be Eddington limited. We discard encounters with $L_\text{peak} < 3 \times 10^{44}~\mathrm{erg~s^{-1}}$, which is comfortably below the errors in Dougie's peak luminosity.

\begin{figure}[tbp]
\epsscale{1.21}
\plotone{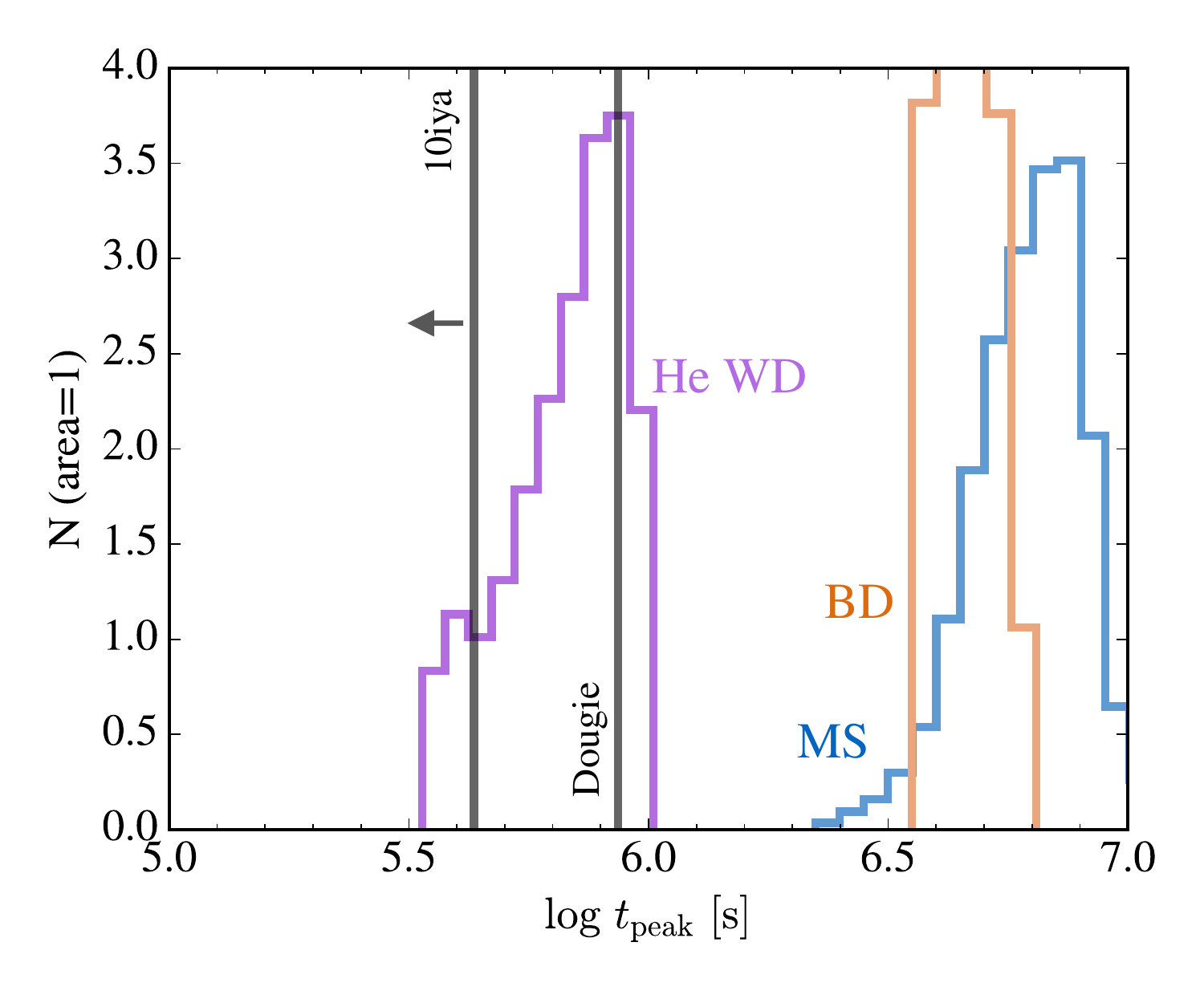}
\caption{
Histograms of peak timescales from the disruption of different types of objects, normalized to area=1, as compared to the peak timescales of two rapidly rising TDE candidates, Dougie and PTF10iya. We include non-He WDs, MS stars (in blue), brown dwarfs (in brown), planets, and the $0.17~M_\sun$ He WD (in purple). We draw from flat distributions in object mass, BH mass, and $\beta$, as described in the text. Peak luminosities are Eddington limited, and we require $L_\text{peak} > 3 \times 10^{44}~\mathrm{erg~s^{-1}}$ in order to reproduce Dougie's peak luminosity.
}
\label{tpeaks_hist}
\end{figure}

In Figure \ref{tpeaks_hist}, we show the outcome of the above exercise. We find that the only objects that satisfy the luminosity requirement are MS stars, BDs, and our prototypical He WD. MS stars and BDs, however, cannot reproduce the rapid timescales of Dougie and 10iya from $\dot M$ alone. Thermal TDEs such as PS1-10jh show a good correspondence between the observed luminosity and the fallback rate \citep{2014ApJ...783...23G}. This simplicity makes the disruption of He WDs an appealing explanation for rapidly rising nuclear transients. 

In order to explain Dougie as a MS star disruption, models require a strong wind component with a functional form that may not directly reflect $\dot M$ \citep{2015ApJ...798...12V}. A wind that carries a significant amount of kinetic and thermal energy may be produced if the accretion rate onto the BH exceeds its Eddington limit \citep{2009MNRAS.400.2070S, 2011MNRAS.410..359L, 2016MNRAS.461..948M, 2016ApJ...830..125J}. While this scenario could explain Dougie and other rapidly rising TDEs such as PTF10iya, their timescales can be naturally explained by the $\dot M$ from He WD disruptions.

We note that \citet{2015ApJ...798...12V} found that Dougie appears offset $\approx3.9$ kpc from the photometric center of its host galaxy. This initially seems to disfavor a TDE interpretation. However, the photometric center of a galaxy is not necessarily its dynamical center. Vink$\acute{\text o}$ et al. also noted that lower-mass off-center BHs are rare yet not unprecedented \citep[e.g.,][]{2008ApJ...683L.119B,2012ApJ...750L..24R}, making the TDE hypothesis tenable.

\subsection{Caveats}\label{subsec:caveats}

Our study focuses on a single example of the disruption of a prototypical $0.17~M_\sun$ He WD. However, as we saw in Section \ref{subsec:properties}, these objects can have a wide range of masses and radii, and the radius evolution even for a single mass is appreciable (see Figure \ref{abundances_burning}). The inclusion of hydrogen-bearing He WDs with a larger range of core masses and envelope masses could potentially explain events with shorter or longer timescales than the prototypical encounters presented here.

In this work, we model the interaction between only a single He WD and a BH. We expect, however, that many He WDs will be in binary systems as they approach the BH, composed of either two He WDs or one He WD and one CO/ONe WD. This suggests that some disruptions of He WDs involve two stars instead of one \citep{2011ApJ...731..128A}. The interaction of the binary with the BH can shift the distribution in binding energy of the debris, and cause the time of peak accretion to occur either earlier or later depending on the sign of the energy shift \citep[a similar effect is seen in disruptions of stars on elliptical orbits;][]{2012EPJWC..3901004H, 2013ApJ...775L...9D}. In extreme cases, this interaction can bind all of the material to the BH (as opposed to just half), allowing the BH to accrete the whole star; alternatively, all of the material can become unbound, preventing any accretion onto the BH. If the binary separation is of order the tidal radius, double tidal disruptions are possible \citep{2015ApJ...805L...4M}. However, our single-star calculations are still applicable for double disruptions, as the hydrodynamics of the disruption are independent for each of the components of the binary. In cases where the outgoing debris streams from the two disrupted stars do not interact with one another, the fallback resulting from a binary disruption can be mimicked by applying simple shifts to the binding energy distribution of the debris of the single-star case.

We do not consider general relativistic effects in our disruption calculations---the gravitational potential of our point mass is purely Newtonian. \citet{2014PhRvD..90f4020C} investigated relativistic effects on the fallback rate of debris. For highly relativistic encounters, they found a more gradual rise and delayed peak of the fallback compared to the Newtonian result. For a $1~M_\sun$, $1~R_\sun$ MS star encounter with a $10^7~M_\sun$ BH, where $r_\text{p}/r_\text{g} \approx 10$, they found a difference in $\dot M_\text{peak}$ of $\approx 18\%$ and a difference in $t_\text{peak}$ of $\approx 10\%$ between Newtonian and relativistic simulations. For a $0.6~M_\sun$ WD encounter with a $10^5~M_\sun$ BH, where $r_\text{p}/r_\text{g} \approx 4.6$, the difference in $\dot M_\text{peak}$ is $\approx 69\%$ and the difference in $t_\text{peak}$ is $\approx 49\%$. 

For the He WD encounters presented in this work, the critical $\beta$ of full disruption has $r_\text{p}/r_\text{g} \approx 12$, and the transition between an envelope-stripping encounter and one penetrating the core occurs at $r_\text{p}/r_\text{g} \approx 19$. Thus, relativistic corrections to our results should be small. In scaling to higher BH masses, however, our errors will increase. However, this will not weaken (and will in fact strengthen) our conclusions regarding the ability of He WDs to achieve the peak timescales of rapidly rising TDE candidates such as Dougie and PTF10iya though $\dot M$ alone, as the relativistic effect is to lengthen the peak fallback timescale.

We use a nested polytrope matched to a MESA profile of the He WD as the initial condition in our disruption calculations. We also track only two fluids---one for the core and one for the envelope---in the simulation, and make the simple choice to model the core as fully helium and the envelope as fully hydrogen. A more realistic treatment might use the MESA profile directly in the disruption calculations, and track the composition of the object more fully. For the particular object used in the simulations in this work, however, the gains in accuracy (aside from composition information) in using the MESA profile directly may be minimal, as the nested polytrope profile is very close to the true profile.

\subsection{Conclusions}

We have modeled the tidal disruption of a new class of object: the low-mass He WD with an extended hydrogen envelope. These objects are a missing link both hydrodynamically and in terms of BH masses probed through prompt tidal disruption flares. In summary, we find that:
\begin{enumerate}
\item Because of their lower density cores and extended envelopes, these objects extend the potential BH masses probed by single-star evolution WDs. In general, their peak fallback timescales will be longer that those of typical WDs and shorter than those of MS stars. 
\item Grazing encounters that strip only the envelope will be hydrogen dominated, and---for a very small amount of mass removed---can provide high and often super-Eddington fallback. 
\item Encounters penetrating the core generally have a fallback rate that is hydrogen-dominated in its rise and helium-dominated in its peak and decline, with relative composition versus time a function of impact parameter. 
\item The typical peak accretion rate of He WD disruptions is a few times larger than that of a typical MS disruption. This likely makes these disruptions observable to larger distances, which would make them a larger fraction of the observed total than suggested by their relative population.
\end{enumerate}
These objects are perhaps the last missing piece of a theoretical tidal disruption menu that includes WDs, MS stars, planets, and evolved stars. Constructing this menu is key to better understanding tidal disruptions. The reader is referred to Figures~\ref{menu}, \ref{mdotpeak_vs_tpeak_scatter}, and \ref{mdots_menu} for a summary of the phase space of the menu.

This work may have particular bearing on two puzzling observational aspects of TDEs that have emerged in the past few years. The first is their rates. There is a great deal of uncertainty in the properties of the nuclear star clusters from which stars are fed into disruptive orbits. Most calculations make standard assumptions of a spherical single-mass nuclear star cluster that feeds stars to the BH by a two-body relaxation-driven random walk in angular momentum space. These calculations predict disruption rates of $\gtrsim 10^{-4}~\mathrm{yr^{-1}}$ per galaxy \citep{1999MNRAS.309..447M, 2004ApJ...600..149W, 2016MNRAS.455..859S}, and are in general in tension with the lower observationally derived rates of roughly $10^{-5}~\mathrm{yr^{-1}}$ \citep[e.g.,][]{2014ApJ...792...53V}. However, there can be several complicating effects---such as secular relaxation, or the presence of a triaxial potential, rings or disks of stars, and/or a second massive body---and there is a lack of understanding of their relative importance in local galaxies. In addition, we need to better understand the mass spectrum of disrupted stars, in particular given mass segregation \citep[e.g.,][]{2016ApJ...819...70M}.

The second puzzling observation is that a significant fraction of tidal disruptions may arise from unique stellar populations. We are learning that tidal disruption flares may occur preferentially in post-starburst galaxies \citep{2014ApJ...793...38A, 2016ApJ...818L..21F}, and that these types of galaxies are overrepresented as TDE hosts. This remains a mystery. Post-starburst galaxies are elliptical-type galaxies that have experienced a star formation burst that has stopped within the past $\sim 1$ Gyr, leaving these galaxies with both old and very young stars.

If only certain types of stars (which are a small fraction of the population) produce prompt flares for BH masses of $\sim 10^6~M_\sun$ due to circularization effects, this could alleviate some of the tension in the observed flaring versus disruption rate. As we have argued in Section \ref{subsec:rates}, the rate of luminous flare production can be distinct from the disruption rate itself. We have shown in Section \ref{sec:menu} and Section \ref{sec:demographics} that different stellar types probe distinct islands of BH mass when we consider prompt flares. This is strong evidence for a connection between stellar population details and the disruption flare rates. The post-starburst galaxy preference may be due to the production of particular stellar species in the nuclei of these galaxies, rather than in the dynamics of their nuclei. We caution, however, that the stellar population of a galaxy as a whole does not necessarily reflect its nuclear population.

In this work, we have argued that to effectively use TDEs to constrain the mass function of BHs, we need to acknowledge that not all disruptions produce luminous flares. Moving forward likely involves understanding the intersection of nuclear region stellar dynamics, stellar populations, and stellar evolution, along with the hydrodynamics of the disruptions themselves. Targeting the observational characteristics of certain TDEs might offer a way to identify BHs at the low end of the supermassive BH mass range.

\acknowledgements
We thank the members of the 2015 Jerusalem Workshop on TDEs for useful comments and discussions. We thank the anonymous referee for constructive comments. The calculations for this research were carried out in part on the UCSC supercomputer Hyades, which is supported by National Science Foundation (award number AST-1229745) and UCSC. M.M. is grateful for support from NASA through Einstein Postdoctoral Fellowship grant number PF6-170155 awarded by the Chandra X-ray Center, which is operated by the Smithsonian Astrophysical Observatory for NASA under contract NAS8-03060. J.G. is grateful for support from Einstein grant number PF3-140108. P.M. is grateful for support from the NSF Graduate Research Fellowship and the Eugene Cota-Robles Graduate Fellowship. E.R.R. is grateful for support from the Packard Foundation and NASA ATP grant NNX14AH37G. This work is also supported by NSF grant AST-1615881. 

\bibliography{refs}
\bibliographystyle{aasjournal}

\end{document}